\documentclass[11pt,a4paper]{article}
\pdfoutput=1

\usepackage{amsmath,amssymb}
\usepackage{slashed}
\usepackage{color}
\usepackage{cite}
\usepackage{jheppub}
\usepackage{braket}

\usepackage{tikz-feynman}
\usepackage{tikz}
\usepackage{amsfonts}
\usetikzlibrary{calc}
\usetikzlibrary{patterns}
\usetikzlibrary{arrows}

\definecolor{emerald}{rgb}{0.31, 0.78, 0.47}

\usepackage{todonotes}
\definecolor{mulberry}{rgb}{0.77, 0.29, 0.55}

\usepackage[normalem]{ulem} %%% for strikethtough

\newcommand\AINote[1]{
 \todo[backgroundcolor=green!20!white,fancyline,
 bordercolor=white]{\scriptsize AI: #1}}
\newcommand\docNote[1]{
 \todo[backgroundcolor=blue!20!white,fancyline,
 bordercolor=white]{DOC: #1}}

\newcommand\ACNote[1]{
 \todo[backgroundcolor=orange!20!white,fancyline,
 bordercolor=white]{\scriptsize AC: #1}}
 
\newcommand{\LCm}{{\scriptscriptstyle -}}
\newcommand{\LCp}{{\scriptscriptstyle +}}

\newcommand{\LCperp}{{\scriptscriptstyle \perp}}
\newcommand{\be}{\begin{equation}}
\newcommand{\ee}{\end{equation}}
\newcommand{\ud}{{\mathrm{d}}}

\renewcommand{\vec}[1]{\boldsymbol{#1}}

\renewcommand{\[}{\begin{equation}\begin{aligned}}
\renewcommand{\]}{\end{aligned}\end{equation}}
\newcommand{\Aop}{\mathbb{A}}
\newcommand{\Fop}{\mathbb{F}}

\title{%Memory and On-Shell Three Point Amplitudes in Minkowskian Quantum Field Theory \\
Large Gauge Effects and the Structure of Amplitudes}
\author[1]{Andrea Cristofoli,}
\author[2]{Asaad Elkhidir,}
\author[2]{Anton Ilderton,}
\author[2,3]{Donal O'Connell}

\affiliation[1]{School of Mathematics and Maxwell Institute for Mathematical Sciences\\
University of Edinburgh, EH9 3FD, United Kingdom}
\affiliation[2]{Higgs Centre, School of Physics and Astronomy, University of Edinburgh, EH9 3FD, UK}
\affiliation[3]{Kavli Institute for Theoretical Physics, University of California, Santa Barbara, CA 93106-4030, USA}

\emailAdd{acristof@ed.ac.uk}
\emailAdd{A.E.H.Elkhidir@sms.ed.ac.uk}
\emailAdd{anton.ilderton@ed.ac.uk}
\emailAdd{donal@ed.ac.uk}

\abstract{
We show that large gauge transformations modify the structure of momentum conservation leading to non-vanishing three-point amplitudes in a simple toy model of a gravitational wave event.
This phenomenon resolves an apparent tension between perturbative scattering amplitude computations and exact methods in field theory.
The tension is resolved to all orders of perturbation theory once 
large gauge effects are included via a modified LSZ prescription; 
if they are omitted, perturbative methods only recover a subset of terms in the full non-perturbative expression.
Although our results are derived in the context of specific examples,
several aspects of our work have analogues in dynamical gravitational scattering processes.
}

\begin{document}

\maketitle

%%%%%%%%%%%%%%%%%%%%%%%%%%%%%%%%%%%%
\section{Introduction}
%%%%%%%%%%%%%%%%%%%%%%%%%%%%%%%%%%%%

The observation of gravitational waves by the LIGO/Virgo/KAGRA collaboration has led to a renaissance in the study of classical aspects of scattering amplitudes~\cite{Neill:2013wsa,Bjerrum-Bohr:2013bxa,Luna:2016due,Luna:2016hge,Luna:2017dtq,Bjerrum-Bohr:2018xdl,Cheung:2018wkq,Kosower:2018adc,Guevara:2018wpp,Bern:2019nnu,Brandhuber:2019qpg,Maybee:2019jus,Guevara:2019fsj,Arkani-Hamed:2019ymq,Cristofoli:2019neg,Bern:2019crd,Bjerrum-Bohr:2019kec,AccettulliHuber:2019jqo,Cristofoli:2020uzm,Bern:2020buy,AccettulliHuber:2020oou,AccettulliHuber:2020dal,Herrmann:2021lqe,Herrmann:2021tct,Bjerrum-Bohr:2021vuf,Bjerrum-Bohr:2021din,Cristofoli:2021vyo,Chiodaroli:2021eug,delaCruz:2021gjp,Bern:2021yeh,Bern:2021xze,Aoude:2022trd,Bern:2022kto,Aoude:2022thd,FebresCordero:2022jts}, see~\cite{Travaglini:2022uwo,Bern:2022wqg,Kosower:2022yvp,Bjerrum-Bohr:2022blt} for recent reviews.
The motivation is precision: future gravitational wave observatories, such as LISA or third generation ground-based detectors, will require a new generation of highly precise gravitational wave templates~\cite{Antonelli:2019ytb,Khalil:2022ylj,Buonanno:2022pgc,Kalogera:2021bya}.
Developments in our understanding of amplitudes have long been motivated by a (successful) drive for precision at particle colliders.
In the application of amplitudes to gravitational wave physics, however, we must also refine our understanding of the classical limit of amplitudes.

Renewed interest in scattering amplitudes, and their connection to classical physics, has been accompanied by a blossoming of ideas about how best to capture the simplifications of the on-shell amplitudes programme while avoiding complications related to separating quantum and classical aspects of amplitudes. 
Effective field theory is a useful tool~\cite{Cheung:2018wkq,Foffa:2019hrb,Foffa:2019yfl,Foffa:2019rdf,Foffa:2019eeb,Levi:2019kgk,Blanchet:2019rjs,Goldberger:2019sya,Aoude:2020onz,Levi:2020kvb,Blumlein:2020znm,Levi:2020uwu,Levi:2020lfn,Blumlein:2020pyo,Foffa:2020nqe,Goldberger:2020fot,Blumlein:2021txj,Foffa:2021pkg,Brandhuber:2021kpo,Brandhuber:2021eyq,Almeida:2021xwn,Kim:2021rfj,Edison:2022cdu,Kim:2022pou,Mandal:2022nty,Kim:2022bwv,Almeida:2022jrv,Mandal:2022ufb}, and has played a crucial role in gravitational wave physics for many years~\cite{Goldberger:2004jt,Goldberger:2005cd,Goldberger:2009qd,Levi:2015msa,Levi:2015uxa,Levi:2015ixa,Levi:2016ofk,Goldberger:2016iau,Foffa:2016rgu,Goldberger:2017frp,Goldberger:2017vcg}, see~\cite{Porto:2016pyg,Levi:2018nxp} for reviews.
We have learned how to take advantage of the fact that amplitudes resum into an exponential form in the classical limit, using eikonal ideas~\cite{tHooft:1987vrq,Amati:1987wq,Amati:1987uf,Jackiw:1991ck,Kabat:1992tb,Ciafaloni:2014esa,Luna:2016idw,Ciafaloni:2018uwe, KoemansCollado:2019ggb,DiVecchia:2019myk,DiVecchia:2019kta,Bern:2020gjj,Parra-Martinez:2020dzs,DiVecchia:2020ymx,DiVecchia:2021ndb,DiVecchia:2021bdo,Heissenberg:2021tzo,Adamo:2021rfq,DiVecchia:2022nna,Adamo:2022rob,DiVecchia:2022piu} as well as WKB-based approximations~\cite{Bern:2021dqo,Kol:2021jjc,Cristofoli:2021jas}.
Since many of the simplifications in amplitudes follow from their relativistic structure, there has been intense recent work on the relativistic ``post-Minkowskian'' effective theory~\cite{Kalin:2020mvi,Kalin:2020fhe,Kalin:2020lmz,Dlapa:2021npj,Dlapa:2021vgp,Kalin:2022hph,Dlapa:2022lmu}.
Worldline quantum field theories enjoy some of the simplifications of both EFT and scattering amplitudes~\cite{Schubert:1996jj,Ahmadiniaz:2016vai,Mogull:2020sak,Jakobsen:2021smu,Jakobsen:2021lvp,Edwards:2021elz,Jakobsen:2021zvh,Jakobsen:2022fcj,Jakobsen:2022psy,Jakobsen:2022zsx}.
It is also worth noting that integration techniques originating from scattering amplitudes have been very useful in gravitational wave theory, e.g. see five-loop and six-loop examples in ~\cite{Bini:2020uiq,Bini:2020rzn}.

Recently a puzzle has arisen relating to static classical modes and very low energy gravitons or photons (we will term these massless quanta \emph{messengers}.)
In a recent paper, Damour~\cite{Damour:2020tta} used a classical (linear response) expression for the radiated angular momentum at order $G^2$ to discuss the contribution of radiation reaction to the order $G^3$ conservative scattering angle.
However, from the perspective of quantum field theory, it is curious that the radiation of angular momentum occurs at order $G^2$ rather than at order $G^3$.
Indeed an analysis of the relevant scattering amplitudes seems to indicate that any radiative process must start at order $G^3$.
There has been a considerable debate about this point in the literature, see for example~\cite{Veneziano:2022zwh,Manohar:2022dea,DiVecchia:2022owy,DiVecchia:2022nna,DiVecchia:2022piu,Gonzo:2022rfk,Bini:2022wrq,Javadinezhad:2022ldc}.

Classically, it is easy to understand the perturbative enhancement of the radiated angular momentum\footnote{We thank Chia-Hsien Shen for explaining this point.} in electrodynamics. 
The angular momentum density is given in terms of a three-dimensional distance vector, and the electric and magnetic fields, by $\vec{r} \times (\vec E \times \vec B)$.
Now the (boosted) Coulomb fields of a static particle fall of as $1/r^2$ and are proportional to a single coupling $e$. 
Thus the static angular momentum density is of order $e^2 / r^3$. 
Integrating this quantity over large spheres leads to a vanishing result as $r\rightarrow \infty$.
In a scattering event, however, there are radiative contributions to the fields of order $e^3 / r$.
Hence, there is a contribution to the angular momentum density of order $e^4 / r^2$ leading to a finite radiation of angular momentum to infinity.
Radiation of linear momentum on the other hand involves integrating $\vec E \times \vec B$, and so the first non-zero contribution during a scattering encounter is of order $e^6$.
Notice that the order $e^4$ term in the radiation of angular momentum involves one static Coulomb field times the radiation field.

From the perspective of scattering amplitudes, we must either declare that the order $e^4$ radiation of angular momentum is somehow not accessible, or else we must understand how to include the static Coulomb mode. 
Ideas for how to do so tend to involve somehow including on-shell three-point amplitudes, describing the coupling of a single messenger to a massive particle, in scattering phenomena.
But, as usually understood, the three-point amplitude only has support on strictly zero-energy messengers in Minkowski space.

This article aims to shine some light on this puzzle by studying what is essentially a toy model, in which we scatter a massive particle and an exact wave solution in electrodynamics and gravity. In principle, it should be possible to think of the exact wave solution either as a coherent state of messengers, or else as a background field~\cite{Kibble:1965zza,Frantz:1965,gavrilov1990qed}.
However, as we shall see, a naive comparison between a perturbative scattering amplitudes-based computation and an all-order background field computation leads to different results, a fact that has been noticed in \cite{Adamo:2022rmp}.
We trace the origin of the discrepancy to a large gauge transformation; in doing so, we will see that this discrepancy is very similar to the puzzle mentioned above.
Tracking the large gauge transformation carefully in the formalism of quantum field theory on a flat background, we find that it leads to a modification of the perturbative amplitudes; for example, altering the support of three-point amplitudes.
Once this is taken into account, there is a complete match between the background computation and the amplitudes.
Along the way, we encounter a memory effect in the linear impulse --- another topic of relevance for gravitational waves.

We begin in the next section by computing the impulse a massive particle suffers in an encounter with a wave. 
We first describe the KMOC~\cite{Kosower:2018adc} approach to computing the impulse, treating the wave as a coherent state and following reference~\cite{Cristofoli:2021vyo}.
We then discuss the background field approach in both electrodynamics and gravity.
As also noticed in \cite{Adamo:2022rmp}, the resulting impulses disagree. In fact the background methods are correct --- and correctly reproduce a classical computation which we briefly review in section~\ref{sec:classicalCheck}.
We turn to the resolution of the apparent conflict in section~\ref{sec:resolve}. 
We first discuss the structure of the coherent states corresponding to the background fields before tracking the large gauge transformation through the LSZ procedure both using background-field and perturbative methods, finally demonstrating a detailed match between the two approaches. We then return to angular momentum in gravitational scattering, before concluding in section~\ref{sec:conclusions}.

%%%%%%%%%%%%%%%%%%%%%%%%%%%%%%%%%%%%
\section{Scattering a particle off a wave}\label{sec:KMOC}
%%%%%%%%%%%%%%%%%%%%%%%%%%%%%%%%%%%%
%
We begin with a discussion of the scattering of a point-like particle with a wave, using the methods of scattering amplitudes.
We will closely follow the KMOC setup~\cite{Kosower:2018adc}, especially as applied to classical wave phenomena in reference~\cite{Cristofoli:2021vyo}.
Our main interest will be in the gravitational scattering of a massive particle from a (classical, gravitational) wave, but as one would expect from the double copy \cite{Kawai:1985xq,Bern:2008qj,Bern:2010ue,Bern:2010yg}, it is frequently helpful to consider the same system in gauge theory. (For our purposes electrodynamics will be sufficient.)
Since we will have occasion to discuss a variety of theories, it is useful to introduce the notation $g$ for a generic perturbative coupling, which could be the electric charge $e$ in electrodynamics or $\kappa = \sqrt{32 \pi G}$ in gravity. 
We set $\hbar=1$ throughout.

%%%%%%%%%%%%%%%%%%%%%%%%%%%%
\subsection{The impulse from amplitudes}
%%%%%%%%%%%%%%%%%%%%%%%%%%%%

From the perspective of amplitudes, it is convenient to describe a classical wave as a coherent state of the relevant messengers (photons in electrodynamics or gravitons in gravity), see~\cite{Ilderton:2017xbj,Cristofoli:2021jas, Britto:2021pud,DiVecchia:2022nna,DiVecchia:2022piu} for applications. 
Coherent states may be defined by the action of the \emph{displacement} operator
$\mathbb{C}(\alpha)$ on the vacuum, where
\begin{equation}\label{CoherentOperator}
\mathbb{C}(\alpha) = \exp \bigg[\int\!\ud\Phi(k) \left( \alpha_\eta^*(k) a_\eta(k) + \alpha_\eta(k) a_\eta^{\dagger}(k) \right)\bigg] \;.
\end{equation}
The creation $a_\eta^{*}(k)$ and annihilation $a_\eta(k)$ operators create or destroy photons or gravitons with helicity $\eta$, while the complex-valued waveshape $\alpha_\eta(k)$ parameterises the specific coherent state\footnote{In our notation, repeated helicity labels are implicitly summed.}. 
The measure factor is defined by
\be
\ud\Phi(k) \equiv \frac{\ud^4 k}{(2\pi)^4} (2\pi) \delta(k^2) \, \theta(k^0) \equiv \hat{\ud}^4 k \, \hat{\delta}(k^2) \,\theta(k^0) \,.
\ee
This is the on-shell phase space measure appropriate for a massless particle; we will also denote the on-shell measure for massive particles as $\ud \Phi(p)\equiv \hat{\ud}^4 p \, \hat{\delta}(p^2-m^2) \,\theta(p^0)$. 
We have further defined hatted differentials and delta functions to absorb factors of $2\pi$.

In terms of the displacement operator, the coherent state itself is
\be\label{displacement-on-vacuum}
\ket{\alpha} \equiv \mathbb{C}(\alpha) \ket{0} \,.
\ee
Because of the exponential in the definition of the displacement operator, this state contains an indefinite number of quanta.
To illustrate the physical interpretation of the coherent state, take an electrodynamic example.
The electromagnetic field operator is
\begin{equation}\label{gaugefieldoperator}
\Aop_\mu(x)= \int\!\ud \Phi(k) \left(  {\varepsilon}_\mu^{\eta}(k) a_\eta(k) e^{-i k \cdot x} + {{\varepsilon}}_\mu^{-\eta}(k) a^{\dag}_{\eta}(k) e^{i k \cdot x} \right) \,.
\end{equation}
It is straightforward to compute the expectation value of this field on the coherent state using the fact that the coherent state is an eigenstate of the annihilation operator:
\be\label{eq:a-alpha}
a_\eta(k) \ket{\alpha} = \alpha_\eta(k) \ket \alpha \,. 
\ee
One finds immediately that $\bra{\alpha} \Aop_\mu(x) \ket{\alpha}$ is given simply by replacing $a_\eta(k)$ in (\ref{gaugefieldoperator}) with $\alpha_\eta(k)$. 
We therefore identify $\varepsilon_\mu^\eta(k) \alpha_\eta(k)$ with the Fourier modes of the classical electromagnetic wave off which we scatter the massive particle. 
Coherent states of gravitons are defined similarly, with the expectation value of the perturbative gravitational field $h_{\mu \nu}(x)$ in the coherent state yielding a classical gravitational wave.

The initial configuration of interest to us contains both a classical wave and a point-like particle.
We therefore consider an initial state of the form
\begin{equation}\label{InitialState}
\ket{\psi, \alpha} = \int\!\ud \Phi(p)\, \varphi(p) e^{i p\cdot z}\ket{p,\alpha} \;,
\end{equation}
where $\ket{p}$ is a single-particle momentum eigenstate of a scalar quantum field, $z$ is a spatial distance and $\ket{p, \alpha} \equiv \ket{p} \ket{\alpha}$.
We will frequently be interested in the classical dynamics of this system,
so we have placed our scalar particle in a (relativistically normalised) wavepacket $\varphi(p)$ so that it is spatially localised.
Obviously there is some uncertainty in both the position and momentum of our particle, but we will
only consider situations where this uncertainty is negligible.
We must also choose coherent states with a classical interpretation: 
this is the case when the expectation value of the number of photons in the wave is sufficiently large.

Having set up our initial state, we now turn to computing
observables.
There are a host of interesting observables to consider, but here we focus on perhaps the simplest:
the impulse, or total change in the momentum of the particle due to its interaction with the wave.
Let $\mathbb{P}_\mu$ be the momentum operator for the massive scalar. Writing the $S$-matrix in terms of the transition matrix $T$ as $S=1+i T$, the particle impulse is defined to be
\[\label{T-TT-KMOC}
\langle\Delta p_\mu\rangle
&= \braket{ \psi,\alpha|S^\dagger [ \mathbb{P}_\mu, S]| \psi,\alpha} \\
&=\langle \psi,\alpha|i[\mathbb{P}_\mu, T]| \psi,\alpha\rangle+\langle \psi,\alpha|T^{\dagger}[\mathbb{P}_\mu, T]|  \psi,\alpha\rangle \,.
\]
In the remaining parts of this section, we will compute this impulse using several different approaches. We will be content to identify the leading order (LO) contribution to the impulse. Working perturbatively in powers of $g$, it is clear that this must come from the first term of \eqref{T-TT-KMOC}. Inserting the initial state (\ref{InitialState}), we have
\begin{equation}\label{4pointImpulse}
    \langle\Delta p_\mu\rangle \simeq \int\!\ud \Phi(p,p')\, \varphi^*(p') \varphi(p) \; e^{i z \cdot (p-p')} i(p'-p)_\mu \bra{p',\alpha} T \ket{p,\alpha}\;,
\end{equation}
where $\ud \Phi(p,p')\equiv \ud \Phi(p)\ud \Phi(p')$.
We now expand the coherent state in terms of number states, using (\ref{CoherentOperator}), which allows us to identify the LO contribution
\begin{equation}\label{4pointImpulse2}
\begin{aligned}
\langle\Delta p^\mu\rangle \simeq i\int\! \ud \Phi(p,p',k,k')\, \varphi^*(p') \varphi(p) \; e^{i z \cdot (p-p')}  (p'-p)^\mu \alpha^*_{\eta'}(k')\alpha_\eta(k) \langle p' , k'^{\eta'}| T | p , k^{\eta} \rangle \;,
\end{aligned}
\end{equation}
in which the matrix element of number states is just the tree-level four-point  amplitude
\begin{equation}
    \langle p' , k'^{\eta'}| T | p , k^{\eta} \rangle = {\hat \delta}^{4}(p'+k'-p-k)  \, \mathcal{A}_4 ( p , k^{\eta} \rightarrow p' , k'^{\eta'}) \;.
\end{equation}
It is clear that the implied LO impulse must be order $g^2$, in both the quantum theory and the classical limit, for any choice of particle momentum and waveshape $\alpha$.

%%%%%%%%%%%%%%%%%%%%%%%%%%%%%%%%%%%%

%%%%%%%%%%%%%%%%%%%%%%%%%%%%%%%%%%%%
\subsection{Solving for the wavefunction}
\label{sec:wavefunction}
%%%%%%%%%%%%%%%%%%%%%%%%%%%%%%%%%%%%
%%%
Scattering amplitudes provide an elegant and systematic method for computing the impulse in perturbation theory. 
However in certain circumstances, and with some caveats, it is possible to identify all-orders contributions to the impulse. 
In this section we explore some examples, determining the impulse resulting from the
scattering of a scalar particle off simple kinds of wave, in both electrodynamics and gravity.
As we shall see, the result is puzzling since it is in apparent contradiction with the impulse
computed in the previous subsection from scattering amplitudes.

%%%%%%%%%%%%%% NEW VERSION HERE.
The basic idea we explore in this subsection is very simple. We will reduce our wave-particle scattering problem to a problem in relativistic quantum mechanics. We will solve exactly for the wavefunction of our particle, and determine the impulse from the wavefunction. We first describe the approach, and then make precise the connection to the coherent state picture above.

Consider the theory described by the action
\be\label{eq:action}
\int \ud^4 x \left( (D_\mu \phi(x) )^\dagger D^\mu \phi(x) - m^2 \phi(x)^\dagger \phi(x)\right) \,,
\ee
where $\phi(x)$ is our scalar quantum field while the covariant derivative is
\[
    D_\mu = \partial_\mu + i e A_\mu(x) \;.
\]
We stress that $A_\mu$, here, is a classical \emph{background} field, chosen to be equal to $\bra{\alpha}\mathbb{A}_\mu(x)\ket{\alpha}$, the expectation value of the quantum field in the coherent state $\ket{\alpha}$. 
Correlation functions, scattering amplitudes and observables in this theory all depend on $A_\mu$, and can be extracted from solutions of the Klein-Gordon (KG) equation in the background $A_\mu$. (The non-relativistic analogue is solving the Schr\"odinger equation.)  To make this clear, consider first a single-particle in-state
\be\label{eq:psi-single-particle}
    \ket{\psi} = \int \ud \Phi(p) \; \varphi(p) \; \ket{p} \,,
\ee
which is nothing but the scalar matter content of the state (\ref{InitialState}). We define the wavefunction associated with this state to be
\be
\label{eq:wf}
    \psi(x) = \bra{0} \phi(x) \ket{\psi} \,,
\ee
where $\phi(x)$ is the Heisenberg field corresponding to the scalar particle.
The wavefunction satisfies the KG equation in the background $A_\mu$ for the simple reason that the quantum field does. 

The explicit form of the wavefunction~\eqref{eq:wf}
depends on the choice of $A_\mu$.
In the asymptotic past, though, the wavefunction simplifies.
Indeed, when the quantum field is free we may write its mode expansion as
\be\label{FreeField}
\phi(x) = \int \ud \Phi(p) \left[ a(p) e^{-i p \cdot x} + b^\dagger(p) e^{i p \cdot x} \right ] \,,
\ee
(where $a(p)$ and $b(p)$ are the mode operators for particles and antiparticles respectively). 
Consequently, in the asymptotic past, the wavefunction is simply
\be
\label{eq:psiStep}
\psi(x) \simeq \int \ud \Phi(p) \, \varphi(p) \, e^{-i p \cdot x} \,.
\ee
In short, $\psi(x)$ is the time-evolved wavefunction of a particle with initial data $\varphi(p)$. 

In the asymptotic future, the form of the mode expansion~\eqref{FreeField} is still valid (albeit with different creation and annihilation operators). Therefore the wavefunction can be written as
\[
\label{eq:psiStepFuture}
\psi(x) \simeq \int \ud \Phi(p') \, \varphi_\textrm{out}(p') \, e^{-i p' \cdot x} 
\]
in the asymptotic future. The outgoing wavepacket $\varphi_\textrm{out}$ is (in general) non-trivially related to the incoming wavepacket $\varphi$.

Of course, it is often very useful to study momentum eigenstates. We therefore define the wavefunction of an eigenstate of momentum $p$ to be
\[
\label{eq:timeEvolvedState}
\psi_p(x) \equiv \braket{0 | \phi(x) |p} \,.
\]
Thus, $\psi_p(x)$ is the solution of the KG equation in the background $A_\mu(x)$ with the boundary condition that
\[
\psi_p(x) \simeq e^{-i p \cdot x}
\]
in the asymptotic past.

Overlaps between different states are basic quantities in quantum field theory. 
Take a second state $\ket{\tilde \psi}$ of the form~\eqref{eq:psi-single-particle}
but defined by a wavepacket $\tilde \varphi(p)$.
Then in momentum space the overlap is the familiar
\[
\label{eq:overlapDef}
\braket{\tilde \psi| \psi} =
\int \ud \Phi(p) \, \tilde \varphi^*(p) \varphi(p) \,.
\]

Given knowledge of the wavefunction $\psi_p(x)$, we can compute the scattering amplitude $\langle p^{\prime}|{S[A]}| p\rangle$ for a particle of initial momentum $p$ to scatter to momentum $p'$, in which we have written $S[A]$ to emphasise that this is the $S$-matrix in the presence of the background $A_\mu$. 
The scattering amplitude is simply the overlap, in the asymptotic future, between $\psi_p(x)$ and an out-state wavefunction of given momentum $p'$.
In position space, we will write this overlap in light-front coordinates $x^\LCp$, $x^\LCm$, $x^i$ where $i= 1,2$, as these will be particularly convenient in the examples that follow. They are defined by the choice of Minkowksi metric
\be
\ud s^2 = 2 \ud x^\LCp \ud x^\LCm - \ud x^i \ud x^i \,.
\ee
In these coordinates, the overlap~\eqref{eq:overlapDef} is given by
\[
\braket{\tilde \psi| \psi} =
\int \hat{\ud}^2 p_\perp \frac{\hat{\ud} p_+}{2 p_+} \theta(p_+) \, \tilde \varphi^*(p) \varphi(p) \,.
\]

To compute the scattering amplitude, we use the solution $\psi_p(x)$ of the KG equation to time-evolve the initial momentum eigenstate to the future.
Notice from equation~\eqref{eq:psiStepFuture} that we can obtain knowledge of the outgoing momentum-space wavepacket from knowledge of $\psi_p(x)$ and vice versa.
Performing the Fourier transform, the scattering amplitude in position space is explicitly
\be\label{eq:2-point-amp}
 \langle p^{\prime}|{S[A]}| p\rangle =  \lim _{x^\LCm \to \infty} \int\!\ud x^\LCp \ud^2 x^\LCperp \, e^{ip'\cdot x}\, 2 i \partial_\LCp \psi_p(x)\;.
\ee
It is easy to write expressions for expectation values of operators in a similar manner.

The approach as described is sufficient for our purposes. Before applying it, let us briefly connect to the perturbative and coherent state methods above. To do so we return to the expectation value of the momentum operator, appearing in (\ref{T-TT-KMOC}): 
\be\label{ExpectationP}
	\langle {\mathbb P}_\mu \rangle \equiv \bra{\psi, \alpha}S^\dagger {\mathbb P}_\mu S \ket{\psi, \alpha} \;.
\ee
We extract the displacement operators as in (\ref{displacement-on-vacuum}), and combine them with the $S$-matrix to obtain
\be\label{eq:Kibble1}
   	\langle {\mathbb P}_\mu \rangle = \bra{\psi}\mathbb{C}^\dagger(\alpha) S^\dagger {\mathbb P}_\mu S \mathbb{C}(\alpha)\ket{\psi}  = \bra{\psi} S^\dagger[A]\, {\mathbb P}_\mu S[A]\, \ket{\psi}  \;.
\ee
Here, we have clarified what we mean by the $S$-matrix on the background $A_\mu$: it is defined~\cite{Kibble:1965zza,PhysRev.139.B1326} to be
\[
S[A]:= \mathbb{C}^\dagger(\alpha) S \mathbb{C}(\alpha) \,.
\]
Applying perturbation theory on the background to equation~\eqref{eq:Kibble1} generates an expansion in powers of $e$, as in vacuum, but where each term is, in principle, exact in $eA_\mu$: 
\be\label{eq:Kibble2}
	\langle \mathbb{P}_\mu \rangle = \int\ud \Phi(p') \, p'_\mu \, |\bra{p'}S[A] \ket{\psi}|^2 + \int\ud \Phi(p',k) \, p'_\mu \, |\bra{p',k}S[A] \ket{\psi}|^2 + \cdots \,.
\ee
The LO contribution to the impulse in this expansion comes from the first term shown. This is the two-point amplitude (on the background), which is of course determined by the KG equation. At the same level of approximation, namely exactly in the quantity $eA$, the time-evolved state
$S \ket{p, \alpha}$ may be written
\[
\label{eq:onlyAlpha}
S \ket{p, \alpha} = \int \ud \Phi(p') \ket{p', \alpha} \braket{p', \alpha| S | p, \alpha} = \int \ud \Phi(p') \ket{p', \alpha} \braket{p'| S[A] | p} \,,
\]
where we we have inserted a complete set of photon states, kept terms of all powers of $eA$, and neglected terms of order $e^2 A$ and so on.
Clearly this returns us to the theory defined by (\ref{eq:action}), where the only amplitude is the two-point on the background. In the examples below we will solve the KG equation \emph{exactly} which means we are, essentially, resumming a whole class of diagrams in QED describing the interaction of the particle with the wave~\cite{Wolkow:1935zz,Kibble:1965zza,Ritus:1985,DiPiazza:2011tq,Fedotov:2022ely}. The neglected higher-order terms, both in (\ref{eq:Kibble2}) and (\ref{eq:onlyAlpha}), correspond to other diagrams which describe the radiation of photons and backreaction on the electromagnetic field, i.e. radiation reaction~\cite{Wistisen:2017pgr,Cole:2017zca,Poder:2017dpw}. The inclusion of these terms, and the extraction of quantum and classical radiation reaction from wave-particle scattering amplitudes is well studied~\cite{Krivitsky:1991vt,Higuchi:2002qc,Ilderton:2013tb,Torgrimsson:2021wcj,Fedotov:2022ely}.

In gravity, the story is absolutely analogous. We only add that we will use the notation $S[h]$ for the $S$-matrix in the presence of a non-trivial gravitational background.
Now we turn to explicit examples. 

\subsubsection*{(i) Electromagnetic ``sandwich'' plane waves}

It is useful to begin with the simplest electromagnetic case: that of plane waves, for which we choose the background electromagnetic field one-form to be
\be\label{eq:A-small-gauge}
    A(x) = A_\mu(x) \, \ud x^\mu = - x^i E_i(x^\LCm) \, \ud x^- \,.
\ee
Notice that the field strength is
\be\label{eq:F-plane-wave}
    F(x) = - E_i(x^\LCm) \, \ud x^i \wedge \ud x^\LCm \,,
\ee
so the electric fields of the wave are given by the functions $E_i(x^\LCm)$. 
It is straightforward to see that all waves of this form solve the Maxwell equations, so we may choose the dependence of the electric fields on $x^\LCm$ freely.
For our purposes, it is very useful to pick $E_i(x^\LCm)$ to be non-vanishing only in a finite
region, for example $0 < x^\LCm < T$ for fixed finite $T$.

\begin{figure}[t!]
    \centering
    \begin{tikzpicture}[scale=0.7, domain=-2*pi:2*pi , rotate=-45,transform shape]
    \filldraw[fill=blue!3!white, draw=black]   (0,0) rectangle (2.8,7);
    \filldraw[fill=blue!3!white, draw=black]   (4.2,0) rectangle (7,7);
    \draw[pattern=horizontal lines light blue]    (2.8,0) rectangle (4.2,7);
    \draw [-> ] (7,7.5) -- (4,7.5);
    \draw [->] (7.5,0) -- (7.5,3);
    \node[font = {\Large\bfseries\sffamily}, black ] at (6,8) { $\boldsymbol{x^- }$};
    \node[font = {\Large\bfseries\sffamily}, rotate around={90:(0,0)}, black ] at (8,1) { $\boldsymbol{x^+ }$};
    \draw [ line width=0.6mm, densely dotted, bend right = 15] (7,0) to (4.2,3.6)  ;
    \draw [ line width=0.6mm,densely dotted](4.2,3.6) -- (2.8,3.9) ;
    \draw [ line width=0.6mm, densely dotted, bend right = 40] (0,7) to  (2.8,3.9) ;
    \node[font = {\Large\bfseries\sffamily}, black ,rotate around={90:(0,0)}] at (2.4,5.7) { $\boldsymbol{x^- = T}$};
    \node[font = {\Large\bfseries\sffamily}, black ,rotate around={90:(0,0)}] at (4.5,5.7) { $\boldsymbol{x^- = 0}$};
    \node (I) at (7,0) {} ;
    \node (J) at (4.2,3.6) {} ;
    \node (K) at (2.8,3.9)  {} ;
    \node (L) at (0,7)  {} ;
    \node  (A) at (1.9,0)  {} ;
    \path (A.east);
    \pgfgetlastxy{\ax}{\ay}
    \node  (B) at (5.1,0)  {} ;
    \path (B.west);
    \pgfgetlastxy{\bx}{\by}
    \pgfmathsetmacro{\mydist}{(\bx-\ax)/28.453}
    \draw[  smooth,samples=200,color=blue,domain=-1:1,shift={($0.5*(A.east)+0.5*(B.west)$)}] plot (\x/2*\mydist,{0.13*sin(\x*1080)/\x});
    \foreach \n in {I,J,K,L}
  \node at (\n)[circle,fill,inner sep=1.5pt]{};
\end{tikzpicture}
    \caption{
    A particle crossing a sandwich plane wave. All particles enter and exit the shaded region at times $x^- = 0$ and $x^-= T$ respectively and have constant velocities outside this region.
    \label{fig:sandwich}
    }
\end{figure}
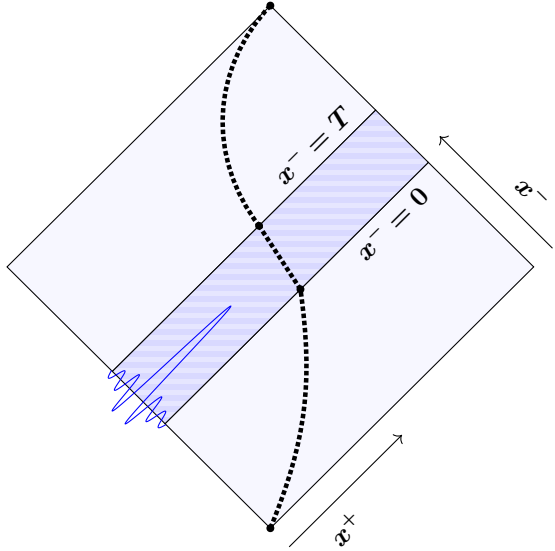

These simple waves have one unphysical property: because the electric field is non-vanishing over an infinite spatial region, the total energy in the wave is infinite.
On the other hand,
they admit a well-defined $S$ matrix (in both gauge theory and gravity)~\cite{Schwinger:1951nm,Gibbons:1975jb,Garriga:1990dp,Adamo:2017nia}.
Waves likes this are sometimes called \emph{sandwich} waves since the spacetime structure
is vacuum in the far past and far future, with only a ``slice'' of non-trivial field strength in the
middle (see figure~\ref{fig:sandwich}).

With these choices, the KG equation to be solved is
\be
    (D^2 + m^2) \psi(x) = 
\partial^2 \psi(x) - 2 i e \, x^i E_i(x^\LCm) \, n_\mu \partial^\mu \psi(x) + m^2 \psi(x) = 0\,,
\ee
where we have introduced the null vector $n_\mu$ defined by $\ud x^\LCm = n_\mu \ud x^\mu$, which will frequently be useful in what follows. We must solve the KG equation subject to the boundary condition that
\be
    \psi_p(x) = e^{-i p \cdot x} \quad\text{for $x^\LCm < 0$.}
\ee
(Note that in light-front coordinates the on-shell condition fixes $p_\LCm = (p_i p_i + m^2) / 2 p_\LCp$.)
As one can easily verify, the solution is the \emph{Volkov wavefunction}~\cite{Wolkow:1935zz}
\begin{equation}\label{VolkovIn}
    \psi_p(x) = \exp \bigg[ -i  p \cdot x + i e \, a(x^\LCm)\cdot x -i \int^{x^\LCm}_{-\infty}\ud s \; \frac{2e \, a (s)\cdot p -e^2 a^2(s)}{2 n \cdot p}  \bigg],
\end{equation}
where $a_\mu \equiv \delta_\mu^i a_i$ is defined in terms of the electric field of the wave as 
\be
\label{eq:potentialDifference}
    a_i(x^{-}) \equiv \int_{-\infty}^{x^{\LCm}} \!\ud s\,  E_i(s) \;. 
\ee
The Volkov wavefunction is an example of a WKB-exact solution of the KG equation.
It is increasingly clear~\cite{Bern:2021dqo,Kol:2021jjc,Cristofoli:2021jas} that solutions of WKB-type are particularly relevant to the application
of scattering amplitudes to classical physics, so the occurrence of a WKB-exact solution here is
a first indication of significant simplifications in the interplay between amplitudes and classical
results.

We introduce here the shorthand 
\[\label{VolPhase}
V_p(x^\LCm) := \frac{2e \, a (x^\LCm)\cdot p -e^2 a^2(x^\LCm)}{2 n \cdot p}
\]
as we will frequently meet this quantity below.

Now that we have an exact solution of the KG equation, it is straightforward to obtain the non-perturbative scattering amplitude. 
Using equation \eqref{eq:2-point-amp}, we find~\cite{Ilderton:2012qe}
\begin{equation}\label{QED2point}
\bra{p^{\prime}}{S[A]} \ket{p} = 
2 p_{+} \hat{\delta}(p_{+}^{\prime}-p_{+}) \hat{\delta}^2(p^{\prime}_\LCperp-p_\LCperp+e a_\LCperp(\infty)) e^{i F_p},
\end{equation}
where the phase $F_p$ is
\be
\begin{split}\label{thephaseF}
    e^{i F_p} &= \lim_{x^\LCm \rightarrow \infty} \exp\bigg[i (p'_\LCm - p_\LCm) x^\LCm - i \int^{x^\LCm}_{0} \! \ud s \, V_p(s) \bigg]  %\; \frac{2e \, a (s)\cdot p -e^2 a^2(s)}{2 n \cdot p} \right] \\
    = \exp\bigg[ i\int_0^\infty \!\ud s\, s\frac{\ud}{\ud s}  V_p(s) \bigg] \;. %\frac{2e \, a (s)\cdot p -e^2 a^2(s)}{2 n \cdot p}\bigg] \,.
\end{split}
\ee
(We used the on-shell condition $p'^2 = m^2$ on the delta function support in equation~\eqref{QED2point}, and the fact that $a(x^\LCm)$ is constant for $x^\LCm > T$, to simplify the $x^\LCm$ dependence.)

The impulse follows from the amplitude~\eqref{QED2point} by inspection of the delta-function constraints and is\footnote{The component $\Delta p_-$ follows from momentum conservation.}
\be
\label{eq:EMimpulse}
    \Delta p_\mu = - e a_\mu(\infty) + n_\mu \frac{2e \, a(\infty)\cdot p - e^2 a^2(\infty)}{2n\cdot p} \;.
\ee
One immediate conclusion is that the impulse is \emph{linear} in the coupling $e$.
This stands in contrast to our previous conclusion that the impulse must be \emph{quadratic} in
the coupling, with corrections involving still higher powers of the coupling.
Something has clearly gone wrong: two different computations of the same physical observable 
lead to results with completely different dependence on the perturbative coupling.

Another indication that something is wrong relates to the perturbative expansion of the 
amplitude~\eqref{QED2point} itself. The amplitude contains non-vanishing terms of order $e$; for
example, such terms follow by expanding the phase $F_p$.
This appears to defy the fact that three-point amplitudes do not go on-shell.

Our goal in the reminder of this paper is to reconcile the results using scattering amplitudes and
the KG equation.
We will see that there is indeed an impulse at order $e$, and that there really is a non-vanishing three-point scattering amplitude in our situation.
To that end, notice that the quantity $a(\infty)$ in equation~\eqref{eq:EMimpulse} is the Fourier zero mode of the electric field strength:
\be
\label{eq:zeromode}
    a_i(\infty) = \int^{\infty}_{-\infty}\!\ud x^\LCm E_i(x^\LCm) \equiv {\tilde E}_i(0) \,,
\ee
or, equivalently, the difference in the electric potential between $x^\LCm = -\infty$ and $x^\LCm=+\infty$. 
This potential difference is ultimately the source of the impulse at order $e$, as we will see explicitly later. 
The enhanced impulse is an example of a memory effect in electrodynamics~\cite{Ilderton:2012qe,Bieri:2013hqa,Susskind:2015hpa,Pasterski:2015zua}.
We will also explore this phenomenon in other other gauges below; we will see that there is a link to large gauge transformations. 
First, we turn to the same basic phenomenon in general relativity.

%%%%%%%
\subsubsection*{(ii) Gravitational plane waves}
%%%%%%%
Gravitational plane waves are conveniently described by the \emph{Brinkmann} gauge metric~\cite{Brinkmann:1925fr}
\begin{equation}\label{metric-Brink}
    \ud s^2=2 \ud x^\LCp \ud x^\LCm - \ud x^a \ud x^a - \kappa H_{a b}(x^\LCm) x^a x^b \ud x^\LCm \ud x^\LCm \;.
\end{equation}
The vacuum equations require $H_{ab}(x^\LCm)$ to be traceless, but place no further restriction on its functional form. 
In this gravitational context, it is helpful to use indices $a, b =1, 2$ for transverse coordinates in Brinkmann gauge.
As above, we focus on sandwich plane waves by taking $H_{ab}$ to be compactly supported on a finite interval $0<x^\LCm<T$, such that there are well-defined in and out regions~\cite{Gibbons:1975jb}. The KG equation to be solved is now
\be\label{KG-GR}
    \left(\frac{1}{\sqrt{-g}} \partial_\mu \sqrt{-g} g^{\mu\nu}\partial_\nu + m^2\right)\psi(x) = \Big(\eta^{\mu\nu}\partial_\mu\partial_\nu +m^2 + \kappa H_{ab}(x^\LCm) x^a x^b \partial_\LCp^2\Big) \psi(x) =0 \;.
\ee
This equation is more complicated than the case of linear plane waves in electromagnetism.
Scalar particles in the gravitational plane wave background were studied by Adamo, Casali, Mason and Nekovar~\cite{Adamo:2017nia}.
Their solution of the KG equation~\eqref{KG-GR}) is expressed in terms of the zweibein $E_{ia}$, labeled by a frame index $i\in\{1,2\}$, obeying
\begin{equation}\label{def-zweibein}
    \ddot{E}_{ia}(x^\LCm)= \kappa H_{a b} E_i^b(x^\LCm) \;,\quad E_i^a(-\infty)=\delta_i^a \;, \quad
    {\dot E}^a_{i}E_{ja} - {\dot E}^a_{j}E_{ia}= 0\;,
\end{equation}
along with its inverse. 
We further define the associated transverse metric $\gamma_{ij}(x^{\LCm}) := E^a_{i} \, E^{\phantom{a}}_{j \, a}$ and the deformation tensor
\[\label{DeformationTensor}
\sigma_{ab}(x^{\LCm}) := \dot{E}^i_a \, E_{i \, b} \;,
\]
which describes geodesic shear and expansion.
Note that indices $a,b\ldots$ are raised and lowered with the flat metric while indices $i,j\ldots$ with $\gamma$. 
In the asymptotic past, the boundary conditions on $E_{ia}$ imply that $\gamma_{ij}\to \delta_{ij}$ and $\sigma_{ab}\to 0$.
The solution itself is $\psi_p(x) = |\gamma(x^\LCm)|^{-1/4} \exp [-i \mathcal{S}(x)]$ where~\cite{Ward:1987ws,Adamo:2017nia,Adamo:2020qru}
\be\label{eq:S-phase-def-GR}
 \mathcal{S}(x) = p_\LCp x^\LCp  + p_i E^i_a x^a +\frac{p_\LCp}{2}\sigma_{ab}x^a x^b +  \frac{1}{2p_\LCp}\int\limits_0^{x^\LCm}\!\ud s  \, \left( m^2 + \gamma^{ij}(s) p_ip_j \right) \;.
\ee
The zweibein, metric and deformation tensor receive, in general, contributions at all orders in the coupling, as does the impulse constructed from the (background) two-point amplitude.
A direct perturbative expansion of this impulse shows that its leading term is $\mathcal{O}(\kappa)$. We thus lose nothing relevant by demonstrating this in a specific case, namely the \emph{impulsive} limit for which $H_{ab}(x^\LCm)\to \delta(x^\LCm) H_{ab}$ with $H$ now a constant matrix. For this field we can solve everything quite explicitly, finding
\begin{equation}\label{zweibein}
    E_{ia} =\delta_{ia}+ \kappa x^{-} \delta_i^b H_{a b} \theta(x^{-}) \;.
\end{equation}
The two-point amplitude becomes
\begin{equation}\label{GravityPhase}
\langle p^{\prime}|S[h]| p\rangle = 2 p_\LCp \hat{\delta} (p_\LCp^{\prime}-p_\LCp)
\frac{2 \pi}{p_\LCp \kappa \sqrt{|\det H}|}
\exp \bigg[-\frac{i}{2 p_\LCp \kappa} (p'-p)_a H^{-1}_{ab}(p'-p)_b \bigg] \;.
\end{equation}
Observe that as $\kappa\to 0$, the amplitude oscillates very quickly.
Defining the limit in a distributional sense, a stationary phase argument
recovers the expected flat-space result, namely an on-shell delta function,
\be
   \lim_{\kappa \to 0} \langle p^{\prime}|S[h]| p\rangle  =
   2 p_\LCp \hat{\delta} (p_\LCp^{\prime}-p_\LCp)\hat{\delta}^2 (p_\LCperp^{\prime}-p_\LCperp) \,.
\ee
This underlines that we are dealing with the $S$-matrix, not the $T$-matrix. From here we calculate the impulse via the expectation value
\be
    \langle \mathbb{P}_\mu \rangle = \int\!\ud\Phi(p')\, p'_\mu |\bra{p'}S\ket{\psi}|^2 \;,
\ee
the evaluation of which amounts to performing a Gaussian integral in $p'$. Unlike the electrodynamic plane wave example above, the position displacement $z$ in the wavepacket now contributes, although the rest of the wavepacket can be dropped under the assumption that it is strongly peaked. We find that the impulse is~\cite{Adamo:2022rmp}
\begin{equation}
\label{eq:GRimpulse}
    \Delta p_\mu = - \kappa h_\mu + n_\mu \frac{2\kappa  h\cdot p -\kappa^2 h\cdot h}{2 n\cdot p} \;, \quad\text{where}\quad h_\mu := p_\LCp\delta_\mu^a H_{ab} z^b.
\end{equation}
This is again of order $\kappa$.
This linear term is a consequence of the velocity memory effect~\cite{Zhang:2017rno,Zhang:2018srn,Shore:2018kmt,Steinbauer:2018iis}. 
Like our earlier impulse, equation~\eqref{eq:EMimpulse}, 
the gravitational impulse~\eqref{eq:GRimpulse} truncates at order $\kappa^2$.
This is simply a consequence of taking the impulsive limit --- for other fields in this class the impulse will contain terms of all higher orders in $\kappa$. 
The impulsive case has the advantage that we can perform the various integrals explicitly, and so serves as a useful example.

Finally, we note that there is a very closely related system in electrodynamics, specifically, with potential
\be\label{A-vortex-small}
    A_\mu(x) = - n_\mu H_{ab}(x^\LCm) x^a x^b \;,
\ee
where $H_{ab}(x^\LCm)$ is again symmetric and traceless, and may be identified with the function appearing in the metric~\eqref{metric-Brink}.
This gauge potential is a generalisation of the vortex solution discussed in reference~\cite{Bialynicki-Birula:2004bvr} and double-copies to the gravitational plane wave~\cite{Ilderton:2018lsf}.
The double copy structure becomes evident upon comparing the KG equation (\ref{KG-GR}) in the gravitational plane wave to the KG equation in this `generalised vortex'; this is
\be\label{KG-vortex}
    \Big(\eta^{\mu\nu} \partial_\mu \partial_\nu + m ^2 - 2 i e H_{ab}(x^\LCm)x^a x^b \partial_\LCp\Big)\psi(x) = 0 \;.
\ee
Notice that the gravitational KG equation~\eqref{KG-GR} contains one additional power of $\partial_\LCp$ relative to the electrodynamic case~\eqref{KG-vortex}, as one would expect from the double copy.

We may trivially obtain, from the gravitational results above, the impulse in the single-copy case of the generalised vortex.  Noting that $p_\LCp$ is conserved in the solution (\ref{eq:S-phase-def-GR}) of the wave equation (\ref{KG-GR}), and comparing with the KG equation (\ref{KG-vortex}), the single copy of the wavefunction, impulse, and other observables are given simply by the replacement $\kappa\to 2e/p_\LCp$.

%%%%%%%%%%%%%%%%%%%%%%%%%%%%%%%%%%%%
\subsection{A classical check}\label{sec:classicalCheck}
%%%%%%%%%%%%%%%%%%%%%%%%%%%%%%%%%%%%
%
We have seen in our examples that the impulse appears at a lower order in $g$ than expected from the KMOC formalism. We therefore perform two checks in the classical limit.

First, we return to the electromagnetic plane wave with field strength (\ref{eq:F-plane-wave}). Classically, we can determine the impulse by solving the coupled Lorentz force and Maxwell equations (for a charge moving and radiating.) 
At low orders we need only study the perturbative solution of the Lorentz force equation
\[
\frac{\ud p_\mu}{\ud \tau} = \frac{e}{m} F_{\mu\nu}(x(\tau)) p^\nu(\tau) \,,
\]
where $p(\tau) = m{\dot x}(\tau)$ is the momentum of the particle and $x(\tau)$ is the position of the particle at proper time $\tau$.
In the case at hand, the electromagnetic field strength tensor is (\ref{eq:F-plane-wave}) and,
focusing on the transverse components of the impulse, we have
\[
\Delta p_i &= - \frac{e}{m} \int_{-\infty}^\infty \ud \tau \, p^-(\tau) E_i(x^-(\tau)) \\
&= -e a_i(\infty) \,.
\]
This is in agreement with our result~\eqref{eq:EMimpulse}.

The gravitational equivalent of this velocity memory effect is easily found. In this case, the equations of motion are given by the geodesic equation
\[
\frac{\ud p^\mu}{\ud \tau}=-\frac{1}{m} \Gamma^\mu_{\alpha\beta}(x(\tau)) p^{\alpha}(\tau)p^{\beta}(\tau) \,.
\]
Focusing again on the transverse components for simplicity, the only nonzero Christoffel symbol needed is
\[
    \Gamma^a_{\LCm\LCm}(x) = - \kappa H_{ab}(x^\LCm) x^b \;.
\]
At lowest order in the gravitational coupling the change in momentum experienced by a particle crossing a gravitational plane wave is
\[
\label{eq:delta-p-grav-1}
\Delta p_{a}= - \frac{p_\LCp^2\kappa}{m} \int_{-\infty}^{\infty}\!\ud\tau \:  H_{a b}(x_0(\tau))x_0^{b}(\tau)  \, ,
\]
in which $x_0^\mu = z^\mu + \tau p^\mu/m$ is the zeroth order (free) solution of the equations of motion.
This leads a non vanishing contribution to the impulse linear in $\kappa$: choosing an impulsive plane wave (\ref{zweibein}) to illustrate and performing the integral over $\tau$ gives the explicit result 
\[\label{eq:delta-p-grav-2}
    \Delta p_a = -\kappa p_\LCp H_{ab}z^b   \,.
\]
in agreement with (\ref{eq:GRimpulse}).

%%%%%%%%%%%%%%%%%%%%%%%%%%%%%%%%%%%%
\section{Reconciliation}\label{sec:resolve}
%%%%%%%%%%%%%%%%%%%%%%%%%%%%%%%%%%%%
%
The background-field calculations above demonstrated a linear contribution to the impulse, consistent with direct classical calculations. The perturbative amplitude calculations we began with, on the other hand, were of order $g^2$, being supported on a four-point amplitude. A linear contribution would have to arise from a three point amplitude, but this vanishes for real momenta by the on-shell condition. We will see that this apparent contradiction is resolved by incorporating into our analysis the effects of \emph{large} gauge transformations which do not vanish (at least sufficiently rapidly) in the asymptotic region. Unlike their small counterparts, large gauge transformations are not mere redundancies in the description of the theory, because they modify boundary conditions in the regions where the transformations do not vanish. In the language of scattering amplitudes, the transformations act nontrivially on the external particle states, and are considered to be genuine symmetries of the $S$-matrix~\cite{Strominger:2013jfa,Strominger:2013lka, He:2014cra, Kapec:2014opa, He:2014laa, He:2015zea, Campiglia:2015kxa,Campiglia:2015qka}. 

We will find below that the introduction of these large gauge effects into our scattering amplitudes modifies their structure leading to the memory effect through a change to the standard LSZ formalism. This provides a simple context in which the connection between memory effects and large gauge transformations~\cite{Strominger:2014pwa,Pasterski:2015tva} is explicitly realized. 
First, we must study the coherent states corresponding to our classical backgrounds.

\subsection{Coherent states}
\label{sec:coherent}

In the perturbative approach, the momentum-space coherent waveshape plays the role of the background field. 
In this section, we will compute the waveshapes and comment on their structure.

It is simplest to begin with the electromagnetic plane wave~\eqref{eq:A-small-gauge}.  One easy way to compute the waveshape is by comparing the Fourier transform of the plane wave's field strength with the field strength expectation value. (The advantage of focusing on comparing field strengths is that they are gauge invariant.) The former is\footnote{We define antisymmetrisation brackets with no associated factor of 2.}
\[
F_{\mu\nu}(x) = \int \hat{\ud}^2 k_\perp \int_{-\infty}^{\infty} \frac{\hat{\ud} k_-}{k_-} k_- n_{[\mu} \delta_{\nu]}^i  \tilde E_i(k_-) \hat \delta^2(k_\perp) e^{-i k \cdot x} \,.
\]
in which $\tilde E_i(k_-)$ is the Fourier transform of the electric field of the wave, while the latter is
\[
\braket{\alpha | \Fop_{\mu\nu}(x) | \alpha}
&= \int \hat{\ud}^2 k_\perp \frac{\hat{\ud} k_-}{k_-} \theta(k_-) \left( i k_{[\mu} {\varepsilon}_{\nu]}^{\eta}(k) \, \alpha_\eta(k) e^{-i k \cdot x} - i k_{[\mu}{{\varepsilon}}_{\nu]}^{-\eta}(k) \, \alpha^*_\eta(k) e^{i k \cdot x} \right) \,,
\]
in which we have written the measure explicitly in lightfront coordinates, as its structure will be important below.

Demanding equality of the field strengths, 
we read off the relation
\[
\label{eq:gettingAlphaStep}
k_- n_{[\mu} \delta_{\nu]}^i  \tilde E_i(k_-) \hat \delta^2(k_\perp) = i k_{[\mu} {\varepsilon}_{\nu]}^{\eta}(k) \, \alpha_\eta(k) \,.
\]
This requirement must be true for some choice of the polarisation vectors and the waveshape.
Without loss of generality we choose linear polarisation vectors
 $\varepsilon_\mu^i = \delta_\mu^i$, so that $i$ becomes the helicity index $\eta = i$.
These vectors are orthogonal to the wave vector $k_\mu = k_- n_\mu$ on the support of the delta function in equation~\eqref{eq:gettingAlphaStep}.
We then read off the waveshape
\[
\label{eq:emWaveShape}
\alpha_i(k) = -i \tilde E_i(k_-) \hat \delta^2(k_\perp) \,.
\]
Specialising for clarity to the impulsive case we have
\[
\label{EMImpulsiveWaveshape}
E_i(x) &= a_i \, \delta(x^-) \,, \\
\alpha_i(k) &= ia_i \, \hat \delta^2(k_\perp) \,.
\]

It is instructive to determine the expectation value of $\Aop(x)$ in this state; the result is
\[
\braket{\alpha | \Aop_\mu(x) |\alpha} = 
\delta_\mu^i a_i \int \frac{\hat \ud k_-}{k_-} i e^{-i k_- x^-} \,.
\]
To define the $k_-$ integral, we must choose a pole prescription. We do so by including an $i\epsilon$ so that
\[
\label{eq:prescription}
    \int \!\frac{\hat \ud k_-}{k_- + i \epsilon} i e^{-i k_- x^-} = \theta (x^-) \,.
\]
This choice is a definition of how to handle the singularity due to the divergent number of zero-energy photons and is reminiscent of the soft dressing suggested in reference~\cite{DiVecchia:2022nna,DiVecchia:2022piu} to incorporate the effects of the static Coulomb mode in scattering observables.

With this pole prescription, we find that our field expectation is
\[\label{eq:emRosen}
\braket{\alpha | \Aop_\mu(x) |\alpha} = 
\delta_\mu^i a_i \, \theta(x^-) \,.
\]
This is not equal to the electromagnetic potential~\eqref{eq:A-small-gauge} we started with!
But that should be no surprise: we determined the waveshape from the field strength, so we can only expect that $\braket{\alpha | \Aop_\mu(x) |\alpha}$ is gauge-equivalent to~\eqref{eq:A-small-gauge}, and indeed it is. 
The gauge-equivalence of this potential with equation~\eqref{eq:A-small-gauge} further justifies our pole prescription~\eqref{eq:prescription}.

Following~\cite{Adamo:2017nia} we refer to~\eqref{eq:A-small-gauge} as \emph{Brinkmann} gauge, and~\eqref{eq:emRosen} as \emph{Rosen} gauge, because of similarities with gauges for the metric of gravitational plane waves. 
The gauge transformation from Brinkmann to Rosen will be important for us.
Defining
\[
\partial_\mu \lambda(x) = A_\mu^\textrm{Brinkmann} - A_\mu^\textrm{Rosen} \,,
\]
we find, in the impulsive case,
\[
\lambda(x) = x^i a_i \, \theta(x^-)\,.
\]
It is important to note that this is a \emph{large} gauge transformation: it does not fall off for large $x$.

It is of course possible to Fourier transform the Brinkmann-gauge potential. We find that
\[
A_\mu^\textrm{Brinkmann} = \int \hat \ud^4 k \, \frac{ \hat \delta(k_+)}{k_-} e^{-i k \cdot x} \frac{\partial}{\partial k_i} \left[ k_\mu \, i a_i \, \hat \delta^2(k_\perp)  \right] \,.
\]
We interpret this expression with reference to equation~\eqref{gaugefieldoperator} as follows. 
The waveshape $\alpha_i(k)$ of equation~\eqref{EMImpulsiveWaveshape} is present as expected.
However, the object playing the role of the polarisation vector is the operator
\[
\frac{\partial}{\partial k_i} \, k_\mu \,.
\]
Operator-valued polarisation vectors are outwith the scope of standard approaches to perturbative scattering amplitudes, though it would be very interesting to explore these more general polarisation objects. 
(See ~\cite{Monteiro:2014cda} for a related application of operator-valued vectors.)
For this reason, the coherent state as we have set it up --- with standard functional polarisation vectors --- only exists in Rosen gauge (or in small gauge transformations thereof). 
Therefore, to use the coherent state and perturbative scattering amplitudes, we are forced to work in Rosen gauge.

More general plane waves, in Rosen gauge, are given by
\[
\braket{\alpha | \Aop_\mu(x) |\alpha} = 
\delta_\mu^i a_i (x^-)  \,,
\]
where the function $a_i(x^-)$ was defined in equation~\eqref{eq:potentialDifference}. The corresponding large gauge transformation is
\[
\label{eq:lambda-general}
\lambda(x) = x^i a_i(x^-) \,.
\]

The situation in gravity is virtually identical, so we will be brief.
The only non-vanishing component of the Riemann tensor for a gravitational plane wave, in Brinkmann coordinates, are
\[\label{Riemann-KerrSchild}
R_{-a-b}&=\kappa H_{ab}(x^{-})\;.
\]
We compare this with the expectation value of the linearised Riemann tensor in a coherent state of gravitons $\ket{\alpha}$:
\[ \label{Riemann-Q-operator}
\bra{\alpha}\mathbb{R}_{\mu \nu \rho \sigma}(x)\ket{\alpha}=\kappa  \operatorname{Re} \int \mathrm{d} \Phi(k) k_{[\mu} \varepsilon_{\nu]}^{\eta}(k) k_{[\rho} \varepsilon_{\sigma]}^{\eta}(k) e^{-i k \cdot x} \alpha_\eta(k) \ ,
\]
with $\alpha_{\eta}(k)$ to be determined by matching (\ref{Riemann-Q-operator}) to (\ref{Riemann-KerrSchild}).  The only essential difference compared to electromagnetism is an additional factor of $k_\LCm$ multiplying the waveshape $\alpha$ on the right hand side. Otherwise, it is again natural to choose, suppressing all indices and helicity labels, $\alpha(k)\sim {\hat \delta}^2(k_\LCperp){\tilde H}(k_\LCm)$.  Considering, for the sake of comparison, an impulsive plane wave, we have
\[\label{wave-GR-imp}
    \alpha(k) \sim \frac{1}{k_{-}} \: \hat{\delta}^{2}(k_{\perp})H \;,
\]
with $H$ constant. Compared to the electromagnetic case (\ref{eq:emWaveShape}), the waveshape now has a pole in $k_{\LCm}$, implying that the number of gravitons is power-law divergent in the IR, compared to logarithmically divergent in QED. We now evaluate the expectation value of a gravitational perturbation in our coherent state,
\[
\bra{\alpha}h_{\mu \nu}(x)\ket{\alpha} &=
\int\!\mathrm{d} \Phi(k) \varepsilon^\eta_{\mu}\varepsilon^\eta_\nu \alpha_{\eta}(k)  e^{-i k \cdot x} + \text{c.c.}
\]
For our choice of waveshape (\ref{wave-GR-imp}) we find, in the impulsive limit and now being careful with indices and polarisation,
\[
\label{eq:RosenAppears}
\bra{\alpha}h_{\mu \nu}(x)\ket{\alpha}= -2 \delta_{\mu}^{i}\delta_{\nu}^{j}\: x^{-} \: \theta(x^{-})H_{ij}\ , 
\]
in which we adopted a similar strategy to (\ref{eq:prescription}). This is not the metric we began with; as in QED, our choice of coherent state waveshape implicitly introduced a coordinate (gauge) transformation, and indeed (\ref{eq:RosenAppears}) is the linearised\footnote{It is interesting to note that some linearised solutions can be generalised to full solutions of the Einstein equations through the existence of a gauge transformation that maps them into Kerr-Schild form~\cite{Harte:2016vwo}.} metric perturbation for a gravitational plane wave in Rosen coordinates~\cite{Adamo:2017nia}.

The momentum-space structure of the Brinkmann gauge metric is very similar to the story in the electrodynamic plane wave.
In particular, it again involves operator-valued polarisation tensors which are outside the scope of our usual methods in perturbative scattering amplitudes.

To understand the physics of our waves in perturbative quantum field theory, then, our programme is as follows. 
We will build on the successful computation we described in the last section using background field methods, but we will perform a gauge transformation/diffeomorphism to Rosen gauge where we have a sensible coherent state.
While changing gauge, we need to be very careful about the large nature of the gauge transformation.
As we will see, once this gauge transformation is performed correctly, we can indeed recover the full scattering physics using the methods of scattering amplitudes.

\subsection{Large gauge transformations and LSZ}
Our goal will be to perform correct perturbative computations in Rosen gauge.
We first study how to reformulate our calculations in Rosen-type gauges, keeping track of large gauge effects.

We focus on the two-point background amplitude (though our arguments apply equally at higher points). 
We require some key formulae from LSZ reduction, see e.g.~the excellent discussion in~\cite[\S 5]{Srednicki:2007qs}. 
Recall that creation and annihilation operators can, at asymptotic times, be written in terms of the field operator $\phi$ by inverting \eqref{FreeField}. 
This allows us to express free mode operators in, for example, the asymptotic future in terms of the full field via
\begin{equation}
    a(p') =\lim_{x^0\to\infty} \int\!\ud^3 {\bf x}\, e^{i p'\cdot x} \, i {\overset{\leftrightarrow}{\partial}}_0  \phi(x) \;.
\end{equation}
The apparent time dependence on the RHS cancels since $\phi(x)$ becomes free in the asymptotic region. Using the corresponding expression for $a^{\dagger}(p)$ in the far past, we immediately arrive at the LSZ reduction formula for the two-point amplitude, 
\begin{equation}\label{LSZReduction}
  \bra{p'} S[A] \ket{p} =
    \lim_{\substack{x^0\to\infty \\ \scriptsize{y^0\to -\infty}}} \bigg[\int\!\ud^3 {\bf x}\, e^{i p'\cdot x} \, i {\overset{\leftrightarrow}{\partial}}_0 \bigg]
    \bigg[-\int\!\ud^3 {\bf y}\, e^{-i p\cdot y} \, i {\overset{\leftrightarrow}{\partial}}_0 \bigg]
    \bra{0} \mathcal{T}\, \phi(x) \phi^\dagger (y) \ket{0}\;,
\end{equation}
in which $\cal T$ is the usual time-ordering symbol. This form of LSZ is not manifestly covariant, and it still contains forward scattering contributions (the ``$1$'' in $S= 1 + i T$), but it is the most useful form for our purposes. Let us now ask what happens when we perform a gauge transformation,
\[
    A_\mu(x) = A'_\mu(x) - \partial_\mu \lambda(x) \;,  \qquad     \phi(x) = e^{i e\lambda(x)} \phi'(x) \;.
\]
$S$-matrix elements, hence (\ref{LSZReduction}), are of course invariant under gauge transformations. 
However, in standard LSZ derivations, it is implicitly assumed that the gauge potential is `small', i.e.~it vanishes asymptotically. 
In situations where the gauge potential is large, we need to go beyond ``standard'' LSZ. To see precisely what changes this entails, consider starting with a small gauge potential and then making a large gauge transformation, choosing a $\lambda(x)$ which vanishes in the far past, but not the far future. Consequently, $A'_\mu$ will be zero in the far past, but become non-zero and pure gauge in the far future: the potential becomes large. 
Inserting the transformation into (\ref{LSZReduction}) we have
\begin{equation}\label{LSZReduction2}
    \begin{aligned}
  \bra{p'} S[A] \ket{p} =
    \lim_{\substack{x^0\to\infty \\ \scriptsize{y^0\to -\infty}}} \bigg[\int\!\ud^3 {\bf x}\, e^{i p'\cdot x} \,
    i \overset{\leftrightarrow}{\partial}_0 \bigg]
    e^{ie\lambda(x)}
    &\bigg[-\int\!\ud^3 {\bf y}\, e^{-i p\cdot y} \,
    i \overset{\leftrightarrow}{\partial}_0\bigg]e^{-ie\lambda(y)} \times \\
    &\qquad\times \bra{0} \mathcal{T}\, \phi'(x) \phi^{\prime\dagger} (y) \ket{0} \;.
    \end{aligned}
\end{equation}
This expression is not symmetric in $x$ and $y$. The additional $y$-dependent phase immediately drops out
because the gauge transformation and potential still vanish in the far past, $y^0\to-\infty$. For the incoming particle LSZ therefore remains unchanged. The situation for outgoing particles is different because $\lambda(x) \neq 0$ for large $x^0$ by hypothesis.

We can simplify this expression substantially.
To do so, first note that the time-ordering operator is inessential in view of the limits on $x^0$ and $y^0$. Inserting a complete set of in-states, we immediately find
\[
\braket{p' | S[A]| p} = 
\lim_{\substack{x^0\to\infty \\ \scriptsize{y^0\to -\infty}}}
\int \ud \Phi(q) 
\bigg[\int\!\ud^3 {\bf x}\, e^{i p'\cdot x} \,
    i \overset{\leftrightarrow}{\partial}_0 \bigg]
    &\bigg[-\int\!\ud^3 {\bf y}\, e^{-i p\cdot y} \,
    i \overset{\leftrightarrow}{\partial}_0\bigg]
e^{ie\lambda(x)} \psi_q(x) e^{i q \cdot y} \,.
\]
We made use of the fact that
\[
\lim_{y^0 \rightarrow -\infty} \braket{q | \phi^\dagger(y) | 0} = e^{i q \cdot y} \,,
\]
since $\ket q$ is an in-state, and of the definition~\eqref{eq:timeEvolvedState}.
It is now a simple matter to show that the amplitude takes the form
\[
\braket{p' | S[A]| p} = 
\lim_{x^0 \rightarrow \infty} 
\int\!\ud^3 {\bf x} \, 
e^{i p'\cdot x}
2i {\partial}_0 \, 
\big( e^{ie\lambda(x)}\psi_p(x) \big)\,.
\]
In lightcone coordinates, the same manipulations result in
\[\label{eq:LSZbackgroundTransformed}
\braket{p' | S[A]| p} = 
\lim_{x^- \rightarrow \infty} 
\int\!\ud x^+ \ud^2 x^\perp \, 
e^{i p'\cdot x}  
2i {\partial}_+ \, 
\big( e^{ie\lambda(x)}\psi_p(x) \big) \,.
\]
Comparing to equation~\eqref{eq:2-point-amp}, we see that the large gauge transformation has simply rephased the final-state wavefunction relative to the initial state, as one would expect.

To illustrate the importance of this modification to the standard LSZ procedure, let us investigate its ramifications in terms of background field calculations, and then perturbative field theory.

\subsection*{Electromagnetic plane waves}
Returning to our potential (\ref{eq:A-small-gauge}), which vanishes in both the asymptotic past and future defined by $x^\LCm\to\pm\infty$, we perform the gauge transformation~\eqref{eq:lambda-general}.
The `primed' potential is thus the Rosen potential
\be\label{eq:Rosen-potential-QED}
    A'_\mu = - x^i E_i(x^\LCm) n_\mu +\partial_\mu (x^i a_i(x^\LCm)) = \delta_\mu^i a_i(x^\LCm) \;,
\ee
which vanishes in the asymptotic past, but not the future, where it goes to the constant value $\delta^i_\mu a_i(\infty)$.  The impact of the same transformation on the scalar wavefunction is simply to remove a phase compared to (\ref{VolkovIn}), hence
\begin{equation}\label{VolkovInRosen}
    \psi^{\prime}(x) = \exp \bigg[ -i  p \cdot x  -i \int^{x^\LCm}_{-\infty}\ud s \; V_p(s)\bigg]\;, 
\end{equation}
and this is, as is easily checked, the solution of the KG equation in the large gauge potential $A'_\mu$. 
Using equation~\eqref{eq:LSZbackgroundTransformed}, we find the `transformed' two-point amplitude to be 
\begin{equation}\label{transformed-but-same}
\begin{split}
  \bra{p'} S {[A]} \ket{p} &=
    \lim_{x^\LCm\to\infty} \int\!\ud^2 x^\LCperp \ud x^\LCp\, e^{i p'\cdot x + i ea_i(x^\LCm)x^i} \, 2 i \partial_\LCp \psi'(x) \\
\end{split}
\end{equation}
in which we passed the phase through the derivative as it is $x^\LCp$-independent; similar simplifications will hold in our other examples. The expression (\ref{transformed-but-same}) straightforwardly recovers equation~\eqref{QED2point}. Suppose, though, that we began from scratch working in the gauge $A'_\mu$, and simply applied `textbook' LSZ, ignoring the fact that the potential does not vanish asymptotically. This would incorrectly yield
\begin{equation}\label{wrong}
\begin{split}
    \lim_{x^\LCm\to\infty} &\int\!\ud^2 x^\LCperp \ud x^\LCp\, e^{i p'\cdot x} \, 2 i \partial_\LCp \phi'(x) \\
    &=2 p_\LCp \hat{\delta}(p'_{+}-p_{+}) \hat{\delta}^2(p'_{\perp}-p_{\perp})
    \lim_{x^\LCm\to\infty} \exp\bigg[ -i \int^{x^\LCm}_{-\infty}\ud s \; V_p(s)\bigg]  \not= \bra{p'}S[A]\ket{p}
    \;. 
\end{split}
\end{equation}
There are two differences compared to (\ref{QED2point}). First, the delta functions have a different support, implying an associated impulse $\Delta p_\mu =0$. Second, the phase is $x^\LCm$-dependent, requiring a prescription for the limit; this is a clear sign that something is wrong, because in asymptotic regions where $E_i=0$ the amplitude is just the KG inner product and should be $x^\LCm$-independent. This underlines that large gauge effects cannot be neglected.

We can consider the same result from another angle. 
In the asymptotic future, a free field in the pure gauge potential $ \delta_\mu^i a_i(\infty)$ with \emph{physical} momentum $p'$ is described by the wavefunction $\phi^\dagger_\text{out}(x) \sim e^{i(p' + ea_i(\infty)) \cdot x}.$
As already pointed out in \cite{Kibble:1965zza,Dinu:2012tj}, it is these `shifted' wavefunctions which appear in the free-field mode decomposition and LSZ reduction formulae when they are expressed in terms of physical quantities. For two-point amplitudes this immediately recovers (\ref{transformed-but-same}). Thus, using either approach, we see that accounting for large gauge effects results in modification of reduction formula and, in particular, a deformation of the momentum conservation delta function in scattering amplitudes.

\subsection*{Gravitational plane waves}
We now turn to gravity. Here similar ideas apply, but the large gauge are now diffeomorphisms, and they act differently. Beginning with the metric (\ref{metric-Brink}), we consider the transformation
\begin{equation}\label{Diffeomorphism}
\begin{aligned}
    y^- &=x^- \,, \\
    y^+ &=x^+ +\frac{1}{2} \sigma_{a b} x^a x^b \,, \\
    y^i &=E_a^i x^a \,,
\end{aligned}
\end{equation}
under which the metric is transformed to `Rosen' form~\cite{EINSTEIN193743}
\begin{equation}
    \ud s^2=2 \ud y^\LCp \ud y^\LCm -\gamma_{i j}(y^\LCm) \ud  y^i \ud y^j.
\end{equation}
The transverse metric $\gamma_{ij}$ reduces to $\delta_{ij}$ in the far past, and generically depends quadratically on $y^\LCm$ in the far future, though it is diffeomorphism-equivalent to the flat metric. Hence we are again dealing with a large gauge transformation. To see the effect of the transformation, we apply (\ref{Diffeomorphism}) to our two-point amplitude (\ref{eq:2-point-amp}). The scalar wavefunction does not pick up an explicit phase as in QED, but rather is evaluated on the new coordinates, so $\psi_p(x) = \psi_p(y)$ where
\begin{equation}\label{GravityWavefunction}
    \psi_p(y)  = | \gamma (y^\LCm)|^{-1/4} \exp \bigg[ - i p_\LCp y^\LCp -i p_i y^i -\frac{i}{2p_\LCp}\int\limits^{y^\LCm}_0\!\ud s \left( m^2 + \gamma^{ij}(s)p_i p_j \right) \bigg].
\end{equation}
The LSZ part of the amplitude changes nontrivially where the diffeomorphism acts on the measure,
\be
    \ud x^\LCp \ud x^\LCperp = \ud y^\LCp \ud y^\LCperp \sqrt{\det \gamma(y^\LCm)} \;,
\ee
and on the exponential factor $e^{ip'\cdot x}$ in the LSZ formula;
\be
    p'_\mu x^\mu = 
p'_\LCm y^\LCm + p'_\LCp y^\LCp + p'_a E_i^a(y^\LCm)y^i +{\frac{p'_\LCp}{4}}{\dot \gamma}_{ij}(y^\LCm)y^i y^j := \Lambda(p',y) \;.
\ee
Consequently, the LSZ formula picks up a determinant and a Gaussian factor in the transverse coordinates, both of which depend nontrivially on $y^\LCm$; the two-point amplitude becomes % 
\be\label{eq:2-point-Rosen}
\begin{split}
\langle p^{\prime}|S[h]| p\rangle &= \lim _{y^\LCm \to \infty} \int\!\ud y^\LCp \ud^2 y^\LCperp \sqrt{\gamma(y^\LCm)}\, e^{i \Lambda(p',y)}\, 2 i {\partial _\LCp} \psi(y) \;.
\end{split}   
\ee
This amplitude should be diffeomorphism invariant (and it is), so let us check again that neglecting large gauge effects yields the wrong answer; in this case we would simply apply the standard LSZ in Rosen gauge, calculating
\be\label{eq:2-point-Rosen-wrong}
\begin{split}
\langle p^{\prime}|S[h]| p\rangle &\not= \lim _{y^\LCm \to \infty} \int\!\ud y^\LCp \ud^2 y^\LCperp e^{i p'\cdot y}\, 2 i {\partial_\LCp} \psi(y)  \propto {\hat \delta}(p'_{+}-p_{+}) \hat{\delta}^2(p^{\prime}_\LCperp-p_\LCperp)
\end{split}   
\ee
for \emph{any} gravitational plane wave, and the implied impulse is $\Delta p_\mu = 0$, in contradiction to (\ref{eq:GRimpulse}). Neglecting the large diffeomorphism would therefore mean missing memory effects, including in particular the `three-point' contribution to the impulse  (\ref{eq:delta-p-grav-1})--(\ref{eq:delta-p-grav-2}).

The single-copy of this discussion yields analogous results for the generalised vortex in electromagnetism. 
Beginning with the potential (\ref{A-vortex-small}) one makes a large gauge transformation with parameter 
\[\label{eq:genVortGT}
\lambda = I_{ab}(x^\LCm)x^a x^b \,,
\]
where $I'_{ab}(x^\LCm) = H_{ab}(x^\LCm)$. The new potential is $A'_\mu = 2 \delta_\mu^a I_{ab}(x^\LCm) x^b$, which is zero in the far past and becomes pure gauge in the far future owing to the symmetry of~$I_{ab}$. As a result LSZ contains a Gaussian integral, like the gravitational plane wave case. Neglecting this shift in LSZ would again lead to an amplitude supported only on forward scattering, and a vanishing impulse. 
One implication of the Gaussian factor, compared to the case of electromagnetic plane waves, is that the scattering amplitude is dependent on the transverse position $z^a$. This leads to the position-dependent single-copy analogue of the impulse (\ref{eq:GRimpulse}).

\subsection{Recovering the non-perturbative amplitude from perturbation theory}

We have seen that large gauge effects must be included in the LSZ prescription to ensure that background-field computations in the Rosen- and Brinkmann-type gauges agree.
We have also seen that our coherent states make most sense in Rosen gauge.
In this section we will show in detail that a perturbative computation in Rosen gauge, including the non-trivial large gauge transformation from Brinkmann to Rosen, leads to complete agreement between perturbative and background-field methods.
One important consequence is that the large gauge transformations modify the structure of momentum conservation, leading to non-vanishing three-point amplitudes in our examples.

First on our order of business is to
recover the $\mathcal{O}(g^N)$ term in the expansion of the (background) two-point amplitude by considering Feynman diagrams with $N$ photon emissions (see e.g.~\cite{Harvey:2009ry,Seipt:2013hda,Adamo:2021hno} for applications of this idea in QED and Yang-Mills).  This means that we want to confirm the equality~\eqref{QED2point} from a purely perturbative scattering-amplitude calculation.
We present the three-point term explicitly here, using the modified LSZ formula \eqref{LSZReduction2}; the remaining terms are discussed in Appendix~\ref{Appendix}.

To access the three-point contribution to $\langle p' , \alpha | S | p , \alpha \rangle$ we expand the coherent states to linear order, and focus on
\begin{equation}\label{Coherent3point}
    \bra{p' , \alpha}S \ket{p , \alpha} \to \int\!\ud\Phi(k) \Big( \alpha_\eta^*(k) \langle p' , k^{\eta}  | S | p \rangle + \alpha_\eta(k) \langle p' |S| p, k^{\eta} \rangle \Big) \,.
\end{equation}
The necessary correlator $\bra{0} \mathcal{T}\, \phi(x) \phi^{\dagger} (y) A_\mu(w) \ket{0}$ is easily computed from the standard Feynman rules:
\begin{equation}\label{eq:correlator}
\begin{split}
    \bra{0} &\mathcal{T} \phi(x) \phi^\dagger(y) {A}_\mu (w) \ket{0} = \\
    &-ie\int \hat{\ud}^4 (p_1,p_2,q)\hat{\delta}^4(p_1 - p_2 - q)
    \frac{i e^{i p_1 \cdot y}}{p_1^2 - m^2 + i\varepsilon} 
     \,\frac{i (p_{1\mu} + p_{2\mu}) e^{-i p_2 \cdot x}}{p_2^2 - m^2 + i\varepsilon}
     \,\frac{-i e^{-i q \cdot w}}{q^2 + i\varepsilon}  \; 
\end{split}
\end{equation}
Applying LSZ to this correlator gives, writing down only the first term on the right hand side of (\ref{Coherent3point}),
\begin{equation}
\begin{aligned}
       \int\!\ud\Phi(k)  \alpha_\eta^*(k) \langle p' , k^\eta  | S | p \rangle = \int\!\ud\Phi(k)  \alpha_\eta^{*}(k)  \lim_{x^\LCm \rightarrow \infty} \bigg[\int\!\ud^3 {\bf x}\, e^{i p'\cdot x}  \,
    i \overset{\leftrightarrow}{\partial}_\LCp \bigg]
    e^{ie\lambda(x)} \\
    \cdots 
     \bra{0} \mathcal{T}\, \phi(x) \phi^{\dagger} (y) A_\mu(w) \ket{0} \,,
     \end{aligned}
\end{equation}
in which the ellipsis stands for the normal LSZ factors for the incoming scalar and the photon, neither of which are modified by the large gauge terms\footnote{There is no issue with a large gauge transformation $\partial_\mu \lambda$ of the photon field because of the usual LSZ contraction with an external on-shell polarisation vector.}. As such the propagators corresponding to $p_1$ and $q$ in (\ref{eq:correlator}) are truncated as usual. What remains is
\begin{equation}
\begin{aligned}
       \int\!\ud\Phi(k)  \alpha_\eta^*(k) & \langle p' , k^\eta  | S | p \rangle  = (- ie)  \lim_{x^\LCm \rightarrow \infty} \bigg[\int\!\ud^3 {\bf x}\, e^{i p'\cdot x + ie \lambda(x)}  \,
    i \overset{\leftrightarrow}{\partial}_\LCp \bigg] \\
    & \times
     \int\!\ud\Phi(k)  \alpha_\eta^*(k) \int \hat{\ud}^4 p_2 \frac{i e^{ -i p_2 \cdot x} \; \varepsilon^{\eta}(k) \cdot (p + p_{2})}{p_2^2 - m^2 + i\varepsilon} \; \hat{\delta}^4(p - p_2 - k)\, .
\end{aligned}
\end{equation}
We perform the integral over $p_{2\LCm}$ by residues, with the appropriate contour specified by the limit $x^\LCm \rightarrow \infty$. The integral over the remaining three components of $p_2$ is then written in a covariant form by introducing the on-shell delta function and is readily evaluated using the overall delta function, yielding
\begin{equation}
\begin{aligned}
    \int\!\ud\Phi(k) & \Big( \alpha_\eta^*(k) \langle p' , k^\eta  | S | p \rangle + \alpha_\eta(k) \langle p' |S| p, k^\eta \rangle \Big) = \\
    & (- {2} i  e) \lim_{x^\LCm \rightarrow \infty} \bigg[\int\!\ud^3 {\bf x}\, e^{i p'\cdot x}  \,
    (p_\LCp + p'_\LCp) \bigg] e^{-i p \cdot x + i a(x^\LCm)\cdot x} \int_{-\infty}^{x^\LCm} \ud s \; \frac{a(s) \cdot p }{2 n \cdot p},
\end{aligned}
\end{equation}
where we have included the second contribution from \eqref{Coherent3point} and made use of the coherent waveshape \ref{eq:emWaveShape} to integrate over photon momenta. Performing the remaining integral we obtain
\begin{equation}
    (-2ie) 2p_+ \hat{\delta}(p_{+}^{\prime}-p_{+}) \hat{\delta}^2(p^{\prime}_\LCperp-p_\LCperp+e a_\LCperp(\infty))
    \lim_{x^\LCm \rightarrow \infty}
    e^{i (p' -p)_- x^-} \int_{-\infty}^{x^\LCm} \ud s \; \frac{a(s) \cdot p}{2 n \cdot p} \,.
\end{equation}
To evaluate the limit, first note that due to the on-shell condition the transverse delta function implies that 
\begin{equation}\label{ShiftedOnShell}
    p'_\LCm - p_\LCm = V_p(\infty)\,. 
\end{equation}
Next, we note that the
exponential factor $e^{i(p'_- - p_-)x^-}$ is present not only in the contribution under discussion, but also in the would-be forward scattering term.
Thus, we have
\begin{equation}
\begin{split}
    \bra{p' \alpha} S \ket{p \alpha}  =   2p_+ &\hat{\delta}(p_{+}^{\prime}-p_{+}) \hat{\delta}^2(p^{\prime}_\LCperp-p_\LCperp+e a_\LCperp(\infty))\\
    &\lim_{x^\LCm \rightarrow \infty}  e^{i (p' -p)_- x^-} \bigg(1 - 2ie\int \ud s \; \frac{a(s) \cdot p}{2 n \cdot p} \bigg).
\end{split}
\end{equation}
The final limit can be expressed more compactly by using \eqref{ShiftedOnShell} and integrating by parts to arrive at the final form:
\begin{equation}\label{BrinkmannLimit}
\begin{split}
    \bra{p' \alpha} S \ket{p \alpha} = 2p_{+}  &\hat{\delta}(p_{+}^{\prime}-p_{+}) \hat{\delta}^2(p^{\prime}_\LCperp-p_\LCperp+e a_\LCperp(\infty)) \\
       & \bigg(1 +i\int_0^\infty \ud s \; s \frac{\ud}{\ud s} \; \bigg(\frac{2e \; a(s)\cdot p}{2 n \cdot p} \bigg)   + \mathcal{O}(e^2)\bigg) \,,
\end{split}
\end{equation}
which is in exact agreement with equation \eqref{QED2point} to this order of precision.

We have now explicitly recovered the three-point contribution from perturbative amplitude methods. By returning to (\ref{Coherent3point}) and retaining higher order terms in the expansion of the coherent state, we could extend our calculation to higher orders in the coupling. Making use of the fact that all the photons in the state are collinear yields significant simplifications, with the ultimate result that the infinite series of perturbative amplitudes exponentiates~\cite{Kibble:1965zza,Frantz:1965}. (We include a demonstration of this fact in appendix~\ref{Appendix}.) 
The result is in agreement with (\ref{QED2point}), confirming the equivalence between coherent state and background field methods, i.e.~$\bra{p'\alpha}S\ket{p\alpha} = \bra{p'}S[A]\ket{p}$.

It follows that the exponent in (\ref{QED2point}) is simply some combination of perturbative scattering amplitudes in vacuum, convolved with the background field profile. 
The fact that the exponent truncates at order $e^2$, rather than containing terms of all orders in the coupling, immediately implies that there must exist an infinite number of relations between perturbative amplitudes on plane wave backgrounds. Let us define `reduced' amplitudes from which we strip momentum-conserving delta functions, so:
\be\label{eq:Expanded}
      \bra{p' \alpha} S \ket{p \alpha}  =  2p_+ \hat{\delta}(p_{+}^{\prime}-p_{+}) \hat{\delta}^2(p^{\prime}_\LCperp-p_\LCperp+e a_\LCperp(\infty)) (1 + {i}\sum\limits_{n=1}^\infty e^n \mathcal{A}_{2+n}) \;.
\ee
These reduced amplitudes can be explicitly computed as follows.
As we have emphasised, the amplitudes receive nonvanishing contributions from both the large gauge transformation and the naive perturbative amplitudes (ie, those obtained from the standard LSZ procedure omitting large gauge effects; in other words, those obtained from Feynman diagrams). 
In particular, the (full) amplitudes are obtained in perturbation theory by applying the modified LSZ formula to a single off-shell current $\mathcal{J}_{n+2}$: the set of Feynman diagrams where only the outgoing massive particle is kept off-shell. 
In Appendix \ref{Appendix} we explicitly demonstrate that the diagrams contributing to $\mathcal{J}_{n+2}$ exponentiate; applying LSZ reduction to the resulting exponentiated form yields equation~\eqref{QED2point} as it should.
Expressed in terms of the coherent state waveshape, this becomes:
\[
    \bra{p' \alpha} S \ket{p \alpha} &= 2p_+ \hat{\delta}(p_{+}^{\prime}-p_{+})  \hat{\delta}^2(p^{\prime}_\LCperp-p_\LCperp+e a_\LCperp(\infty))  \\
    &\times \exp \bigg[  2ie \int\!\ud \Phi(k) k_\LCm\left  ( \alpha_\eta(k)\;  {\varepsilon}^{\eta}(k) \cdot p \; \hat{\delta}'(2p \cdot k) \right)  \\
    & ie^2\int\!\ud \Phi(k,k') (k'_{\LCm} - k_{\LCm}) \left( \alpha_\eta(k) \alpha_{\eta'}(k') {\varepsilon}_{\eta}(k) \cdot {\varepsilon}_{\eta'}(k') \hat{\delta}'(2p \cdot (k'-k)) \right) \bigg] \,. 
\]
We see that the reduced amplitudes differ from the naive amplitudes obtained from the normal LSZ prescription. Specifically, we note the unusual appearance of the derivative of the on-shell delta function. 

The reduced amplitudes can also be obtained
by expanding the exponential $\exp i F_p$ in (\ref{thephaseF}). Equating the result with (\ref{eq:Expanded}) the leading terms are clearly
\begin{equation}
    \mathcal{A}_3 = \int_0^\infty \!\ud s\, s\frac{\ud}{\ud s} \frac{a(s) \cdot p}{n \cdot p}
    \;,
    \qquad
    \mathcal{A}_4 = \frac{i}{2}\mathcal{A}^2_3 -  \int_0^\infty \!\ud s\, s\frac{\ud}{\ud s} \frac{e^2 a^2(s)}{2 n \cdot p} \;.
\end{equation}
We can `invert' these expressions in order to rewrite the two points amplitude on a background in terms of perturbative three- and four-point amplitudes:
\be\label{eq:volkov-as-amps-2}
      \bra{p' \alpha} S \ket{p \alpha}  =  2p_+ \hat{\delta}(p_{+}^{\prime}-p_{+}) \hat{\delta}^2(p^{\prime}_\LCperp-p_\LCperp+e a_\LCperp(\infty)) \exp\bigg[ {i} e\mathcal{A}_3 + {i} e^2 \mathcal{A}_4 {+} e^2\frac12 {\mathcal A}_3^2\bigg] \;.
\ee
Notably, it follows that that we can express \emph{all} higher point amplitudes in terms of $\mathcal{A}_3$ and $\mathcal{A}_4$ by expanding the exponential in this form, obtaining for example
\be\label{eq:volkov-as-amps-1}
\begin{split}
    {\cal A}_5 &= {i}{\cal A}_3{\cal A}_4 { + } \frac{1}{3} {\cal A}_3^{{3}} \;, \\
    {\cal A}_6 &= \frac{{i}}{2} {\cal A}_4^2 {+} \frac{{i}}{12} {\cal A}_3^4 \;, \\
    {\cal A}_7 &= {-} \frac{{\cal A}_3 {\cal A}_4^2}{2} {+} \frac{{i}{ \cal A}_3^3 {\cal A}_4}{3} + \frac{{\cal A}_3^5}{20} \;,
\end{split}
\ee
and so on --- all higher-point vacuum amplitudes factorise. 
From a purely perturbative approach, none of these relations are obvious from the start, but they are all implied by the exponential form of the Volkov solution, and the fact the exponent truncates at second order in the coupling. 
(The situations in gravity and for the electromagnetic vortex is different -- the exponent is a Laurent series in $g$.)
It is now clear that the three-point contribution is essential in order to recover the full result. For example, without the three-point contribution all amplitudes with an odd number of legs, $\mathcal{A}_{2n+1}$, would be found incorrectly to be zero.  
Hence the nonvanishing 3-point amplitude is necessary to obtain the recursive relationships in \eqref{eq:volkov-as-amps-1}.

\subsection{Recovering the impulse}

Earlier we encountered what seemed like a problem: perturbative methods led to an incorrect impulse.
Now that we understand the origin of the problem --- namely that we were neglecting an important large gauge transformation --- we can correctly recover the impulse using perturbative techniques.
We first use the resummed expression to find a convenient form of the final state in a classical scattering event, reminiscent of the eikonal context (see reference~\cite{Cristofoli:2021jas} for a recent discussion.) 
The initial state~\eqref{InitialState} depends on a wavepacket $\varphi(p)$, 
which is sharply-peaked at an initial classical momentum $p^\textrm{initial}_\mu$. Inserting a complete set of states and retaining only dominant terms as in equation~\eqref{eq:onlyAlpha}, the final state is
\[
S \ket{\psi, \alpha} = \int \ud \Phi(p',p)
\, \varphi(p) \, \ket{p', \alpha} \bra{p', \alpha} S \ket{p, \alpha} \,.
\]
Since the resummed amplitude~\eqref{eq:volkov-as-amps-2}
 contains three delta functions, we can integrate over the phase-space of $p$ to find~\cite{Adamo:2022rmp}
\[
\label{eq:finalState}
S \ket{\psi, \alpha} = \int \ud \Phi(p')
\, \varphi(p(p')) \, \ket{p', \alpha} e^{i F_{p(p')}} \,,
\]
where now the quantity $p(p')$ is defined to be
\[
p_\mu(p') = p'_\mu
+ e a_\mu(\infty) - n_\mu \frac{2e \, a(\infty)\cdot p - e^2 a^2(\infty)}{2n\cdot p} \,.
\]
(Note the various sign changes compared to (\ref{eq:EMimpulse}), and see~\cite{Dinu:2012tj} for related discussions.)
With this form of the final state, it is straightforward to compute outgoing observables.
For example, the final momentum is
\[
\braket{\psi, \alpha| S^\dagger \mathbb{P}_\mu S | \psi, \alpha}
&=
\int \ud \Phi(p',p'') \, \varphi^*(p(p'')) \varphi(p(p')) \, \braket{p'', \alpha| \mathbb{P}_\mu | p', \alpha} e^{i \left(F_{p(p')} - F_{p(p'')}\right)} \\
&=
\int \ud \Phi(p') \, \varphi^*(p(p')) \varphi(p(p')) \,  p'_\mu \,.
\]
Since the wavepackets are steeply-peaked at $p(p') = p^\textrm{initial}$, the integration is trivial, and recovers our previous expression~\eqref{eq:EMimpulse} for the impulse.

The situation in the generalised electromagnetic vortex~\eqref{A-vortex-small} is similar though more complicated.
Once LSZ is correctly implemented, the perturbative amplitudes exponentiate and satisfy
\[
\braket{p', \alpha| S | p, \alpha} = \braket{p'|S[A]| p} \,,
\]
where now the background $A$ is the vortex.
Since the relevant large gauge transformation~\eqref{eq:genVortGT} is quadratic in position, the resummed amplitude involves Gaussians instead of the naive delta functions.
The final state $S\ket{\psi}$ can again be simplified using
\[
\label{eq:basicStory}
S \ket{p, \alpha} &= \int \ud\Phi(p') \ket{p', \alpha} \braket{p', \alpha|S|p, \alpha} \\
&= \int \ud\Phi(p') \ket{p', \alpha} \braket{p'| S [A]|p} \,.
\]
The story for the gravitational plane wave is essentially the same as for the generalised vortex, as one would expect from the double copy.
The perturbative amplitudes again exponentiate into $\braket{p'| S[h]| p}$ --- explicitly given for the impulsive case in equation~\eqref{GravityPhase} --- and again one encounters a Gaussian function where naively a delta function would be expected.
Equation~\eqref{eq:basicStory} holds with the background $S$-matrix set to $S[h]$.

\subsection{Angular momentum}
In this section we consider the effect of the large diffeomorphism \eqref{Diffeomorphism} on the change of angular momentum experienced by a pair of particles crossing a gravitational plane wave.

Consider two particles which, in the far past $x^\LCm\to-\infty$, are situated symmetrically around the line $z^+=0$, at positions $(z^\LCp , z^i)$ and $(-z^\LCp , -z^i)$. 
We will calculate the change of angular momentum experienced by each of these particles. 
To this end we require the angular momentum operator $\mathbb{J}_{\mu\nu}$ which, expressed in terms of creation/annihilation operators of the free field, is\footnote{Our antisymmetrization brackets are defined without a factor of $2$, but we do include a factor of $2$ in our double-sided derivative.}~\cite{Manohar:2022dea,DiVecchia:2022owy}
\begin{equation}
    \mathbb{J}_{\mu \nu} =  \int\!\ud \Phi(p)\, a^{\dagger}(p) i p_{[\mu} \frac{\stackrel{\leftrightarrow}{\partial}}{\partial p^{\nu ]}} a(p) \;.
\end{equation}
To obtain the expectation value of $\mathbb{J}_{\mu \nu}$ in the final state, we express the final state in the form \eqref{eq:finalState} where, in the impulsive limit, the momentum shift is
given by equation~\eqref{eq:GRimpulse}. 
As in the QED case, the final state $S \ket{\psi, \alpha}$ acquires a phase due to the exponentiation of soft gravitons. This phase cancels in the overlap 
with the conjugate state and thus does not contribute to the expectation value $\bra{\psi,\alpha} S^\dagger  \mathbb{J}_{\mu \nu} S \ket{\psi,\alpha}$. The change in angular momentum is therefore obtained solely from the deformed wavefunction. Taking for example the transverse components of angular momentum we find 
\begin{align}
\label{eq:expt-J}
    \bra{\psi,\alpha} S^\dagger  \mathbb{J}_{ij} S \ket{\psi,\alpha} &= \int \ud \Phi(p',p'') \, \varphi^*(p(p'')) \varphi(p(p')) \, \braket{p'', \alpha| \mathbb{J}_{ij} | p', \alpha} e^{i \left(F_{p(p')} - F_{p(p'')}\right)}  \nonumber \\
        &=  -\int \ud \Phi(p') \, \varphi^*(p(p')) \varphi(p(p')) p'_{[ i} z_{j]} \nonumber \;.
    \end{align}
Integrating out the wavepacket, in the limit that it is sharply peaked as usual, has the effect of replacing $p_\mu$ with the on-shell value implied by the argument of the wavefunction. The change in angular momentum is given by subtracting, from this result, the initial angular momentum, $\bra{\psi}  {\mathbb J}_{ij}\ket{\psi}$, so that
\begin{equation}\label{delta-J}
    \Delta  J_{i j} = -  \kappa p_\LCp H_{ab}z^b \delta^a_{[i}z_{j]} \;. 
\end{equation}
The change in angular momentum for the system of the two particles described above is \emph{twice} the contribution in (\ref{delta-J}); the quadratic dependence of $\Delta J_{ij}$ on the transverse coordinates $z^i$ means that the two contributions sum up, rather than cancel. This result is $\mathcal{O}(\kappa)$, lower than expected for the change in mechanical angular momentum in ordinary perturbative scattering. This is due to fact that there is an $\mathcal{O}(\kappa)$ contribution in the impulse, which we know results from the action of the large diffeomorphism on the amplitude. The same result holds for the electromagnetic vortex, upon making the `single copy' replacement $\kappa\to 2e/p_\LCp$ as before.

This low-order $\Delta J_{ij}$ is reminiscent of the situation discussed in~\cite{Veneziano:2022zwh}, where the appearance of radiated angular momentum at $\mathcal{O}(G^2)$~\cite{Damour:2020tta}, as opposed to the expected $\mathcal{O}(G^3)$, was attributed to the ambiguity of choosing a BMS gauge. 
In our context, we also encounter a parametrically enhanced change of angular momentum. 
We have concretely connected this phenomenon to large gauge effects, which alter the basic structure of naive perturbative amplitudes.
It will be interesting to explore the connection between LSZ, large gauge effects, and scattering amplitudes further in future.

%%%%%%%%%%%%%%%%%%%%%%%%%%%%%%%%%%%%
\section{Conclusions}\label{sec:conclusions}
%%%%%%%%%%%%%%%%%%%%%%%%%%%%%%%%%%%%
%
We have examined the role of large gauge transformations in wave-particle scattering, in both QED and gravity. 
We began by highlighting an apparent tension noticed in \cite{Adamo:2022rmp} between perturbative scattering amplitude computations, exact methods in field theory, and direct classical calculations, in which the same radiative observables were found to be $\mathcal{O}(g)$ or $\mathcal{O}(g^2)$ depending on the approach taken. 
By studying the properties of the coherent states which describe the wave in the scattering process, we identified the presence of large gauge contributions. We showed that these modify the structure of momentum conservation in perturbative methods, leading to non-vanishing three-point amplitude contributions. Once these are taken into account, there is a complete match between all methods of computation, and perturbative methods recover the classical and background field results.

Although our work focused on a selection of exact solutions in electromagnetism and gravity, there are a number of provocative similarities between our cases and the situation in gravitational scattering.
One example which we highlighted is that the change in angular momentum of a particle, as it passes through a wave, can be of lower order than one would expect from perturbation theory.
We also saw that the coherent waveshape $\alpha$ plays an important role in the all-order structure of the amplitude.
In particular, there can be a deformation of the basic structure of momentum conservation we naively expect in scattering amplitudes.
As we saw in equation~\eqref{eq:zeromode}, this dressing is connected to zero-energy messengers.
In gravitational scattering, a similar dressing~\cite{DiVecchia:2022owy} correctly captures the radiated angular momentum~\cite{DiVecchia:2022owy,DiVecchia:2022piu}, including the BMS dependence of the $O(G^2)$ angular momentum loss~\cite{Veneziano:2022zwh}.

While most of the interest in the classical particle scattering is in the gravitational case, it is worth noting that there is also an enhancement in the loss of angular momentum in electrodynamics. 
We have seen a similar parametric enhancement occurs in the background of the generalised electromagnetic vortex, which double copies to the gravitational plane waves we have considered.
Unlike the situation in gravity, where angular momentum is subject to a BMS ambiguity, $\Delta J_{ij}$ is gauge invariant in electrodynamics. Thus, the issue of angular momentum loss in QED is in some respects sharper than in gravity~\cite{Bonga:2018gzr}.
It would be interesting to understand this situation more clearly, particularly from the perspective of the double copy.

Our waves are idealised toy models, and it would in future be useful to extend our analysis to different wave solutions, such as generic pp-waves.
Still, plane waves have several useful properties: it is possible to find exact solutions of the KG equation in their presence, and they provide a toy model in which to understand caustics, strong lensing, and tails effects~\cite{Harte:2012uw,Harte:2012jg,Harte:2013dba,Harte:2015ila,Flanagan:2018yzh}. 
More realistic waves would be damped in all spatial directions,
and are easily dealt with perturbatively --  a non-perturbative treatment is however challenging due to the difficulty of solving the KG equation, see e.g.~\cite{Friedlander:2010eqa} for a review.

The exact wavefunctions that played an important role in our work are also the solutions that one would find using the WKB approximation --- in other words, the WKB approximation happens to be exact in the circumstances of interest to us.
This connects our work again to an important set of ideas in the literature relating scattering amplitudes to classical physics.
The fact that perturbative amplitudes, in the classical regime, can be resummed into an exponential form is the key idea of the eikonal approximation~\cite{Amati:1987wq,Amati:1987uf,Ciafaloni:2014esa,Luna:2016idw,Ciafaloni:2018uwe, KoemansCollado:2019ggb,DiVecchia:2019myk,DiVecchia:2019kta,Bern:2020gjj,Parra-Martinez:2020dzs,DiVecchia:2020ymx,DiVecchia:2021ndb,DiVecchia:2021bdo,Heissenberg:2021tzo,Adamo:2021rfq,DiVecchia:2022nna,Adamo:2022rob,DiVecchia:2022piu}.
A closely-related resummation connects scattering amplitudes to the classical radial action and the Hamilton-Jacobi equation~\cite{Bern:2021dqo,Kol:2021jjc,Cristofoli:2021jas}.
It may be fruitful to examine whether the perspective we gained from plane wave calculations can be applied to the more complicated context of BMS transformations and the ambiguity in radiated angular momentum building on approximate WKB methods. 

Another interesting direction is to understand the relations between on-shell amplitudes on flat spacetimes and $n$-point amplitudes on curved backgrounds \cite{Adamo:2020yzi,Adamo:2022mev} in more depth. 
For the case of plane wave backgrounds \cite{Adamo:2020qru}, it has recently been shown that these quantities naturally reproduce gravitational self-force results.
The self-force technique is an expansion in energy ratios and therefore naturally resums the post-Minkowskian approximation to all orders~\cite{Adamo:2022qci}. 
This could open the way to the application of on-shell methods to alternative perturbative schemes like the self-force programme \cite{Poisson:2003nc,Harte:2017hfp,Harte:2018iim}.

\acknowledgments

We thank Tim Adamo, Julio Parra Martinez, Rodolfo Russo, Chia-Hsien Shen, Mao Zeng.
This research was supported in part by the National Science Foundation under Grant No. NSF PHY-1748958.
AE is sponsored by a Higgs Fellowship.
DOC is supported by the U.K. Science and Technology Facility Council (STFC) grant ST/P000630/1. AC is
supported by the Leverhulme Trust (RPG-2020-386).
For the purpose of open access, the author has applied a Creative Commons Attribution (CC BY) licence to any Author Accepted Manuscript version arising from this submission. %\docNote{Andrea funding?}

\appendix
\section{Exponentiation}\label{Appendix}
In what follows we will illustrate the exponentiation of the amplitude $\bra{p' \alpha} S \ket{p \alpha}$ in a plane wave background \cite{Kibble:1965zza, Frantz:1965} . To do this, we first consider the diagrams contributing to order $\mathcal{O}(e^N)$, keeping all but the outgoing massive leg on shell, we then treat the LSZ reduction of the outgoing particle separately. We denote the contributions of these diagrams at $\mathcal{O}(e^N)$ by $\mathcal{J}_{2+N}$, so that the $S$-matrix is obtained by applying the LSZ truncation to the outgoing leg:
\begin{equation}\label{LSZFinalLeg}
    \mathcal{A}_{2+N} =  \lim_{x_\LCm \rightarrow \infty} \bigg[\int\!\ud^3 {\bf x}\, e^{ip'\cdot x + ie \lambda(x)}  \,i \overset{\leftrightarrow}{\partial}_\LCp \bigg]  \mathcal{J}_{2+N}
\end{equation}

In the limit $x^\LCm \rightarrow \infty$, the Feynman propagator on the outgoing leg reduces to a retarded propagator. In this limit, we can write the contribution of $\mathcal{J}_{2+N}$ as
\begingroup 
    \tikzset{every picture/.style={
        ,scale=2.2, transform shape
    }}
    \begin{equation}
\begin{aligned}
\mathcal{J}_{2+n} &= \int \hat{d}^4 p_2 \theta(p_+) \hat{\delta}( p_2^2 - m^2)  e^{-i p_2 \cdot x} \hat{\delta}^4 \left(p_2 - p - \sum_{i=1}^N q_i \right). \\
 & \times \left(
\begin{gathered}
\begin{tikzpicture}[scale = 0.6,]
  \begin{feynman}[scale = 0.6,]
    \vertex (i); 
    \vertex [right=of i] (i1);
    \vertex [above=of i1] (f1);
    \vertex [above right= 1.4 of i1] (r1)  {\( q_2 \)};
    \vertex [above left = 1.4 of i1] (l1)  {\(q_1 \)};
    \vertex [right=of i1] (i2) ; 
    \vertex [above=of i2] (f2) ;
    \vertex [right=of i2] (i3)  ; 
    \vertex [above=of i3] (f3) ;
    \node[right = 0.01 cm of i2] (a)  {\(\cdots\)};
    \vertex (j)  [right= 0.3 cm of a] ; 
    \vertex [right=of j] (j1);
    \vertex [above=of j1] (f1);
    \vertex [above right=1.4 of j1] (r2)  {\( q_{2m} \)};
    \vertex [above left = 1.4 of j1] (l2)  {\(q_{2m-1} \)};
    \vertex [right=of j1] (j2) ; 
    \vertex [above=of j2] (f2) ;
    \vertex [right=of j2] (j3)  ; 
    \vertex [above=of j3] (f3) ;
    \node [right = 0.01 cm of j2] (b)  {\( \cdots \)};
    \vertex (k) [right= 0.3 cm of b]; 
    \vertex [right=of k] (k1);
    \vertex [above= 1cm of k1] (g1) {\( q_{N-1}\)};
    \vertex [right=of k1] (k2) ; 
    \vertex [above = 1cm of k2] (g2) {\( q_N \)} ;
    \vertex [right=of k2] (k3) ; 
    \vertex [above=of k3] (g3);
    \diagram* {
    (i) -- [fermion , edge label' = $p$, near start] (i1) -- [edge label' = $p+\Tilde{q}_1$,near end](i2),
    (i1) -- [photon] (r1) ,  
    (i1) -- [photon] (l1) , 
    (j) -- (j1) -- (j2) ,
    (j1) -- [photon] (r2) ,  
    (j1) -- [photon] (l2) , 
    (k) -- (k1) -- [edge label' = $p + q_{n-1}$]  (k2) -- [ opacity=0.2 , ultra thick ] (k3) ,
    (k1) -- [photon] (g1) ,  
    (k2) -- [photon] (g2) , 
    };
  \end{feynman}
\end{tikzpicture}
\end{gathered}
\right) + \text{permutations ,}
\end{aligned}
\end{equation}
\endgroup
where we have shown the diagram in $\mathcal{J}_{2+N}$ consisting of $m$ four-point vertices and $N-2m$ 3-point vertices, with all four-point vertices placed before all three-point vertices. We have also isolated the contribution of the outgoing leg into the prefactor multiplying the diagram. As we will see, the full result can be obtained by evaluating this diagram and subsequently summing over permutations. Evaluating the diagram in this arrangement and making use of the delta function yields 
\begin{equation}
\begin{aligned}
     \hat \delta(2p \cdot (\sum_{i=1}^{m} \tilde{q_i} + \sum_{j=2m+1}^{N} q_i ))\frac{( \varepsilon \cdot \varepsilon)^m (\epsilon \cdot p)^{N-2m}}{(2p \cdot \tilde{q}_1) \cdots (2p \cdot (\sum_{i=1}^m \tilde{q}_i)(2p \cdot (\sum \tilde{q}_i + q_{2m+1})) \cdots (2p \cdot (\sum_{i=1}^{N-1} q_i)}\\
     \times \exp \left[ -i(p+(\sum_{i=1}^{N} q_i)) \cdot x \right] ,
\end{aligned}
\end{equation}
where we have made use of the lightfront gauge condition $\varepsilon(q_i) \cdot q_j=0$ to write the contribution of three-point vertices as $\varepsilon(q_i) \cdot p$ regardless of where the vertex is placed in the diagram, and used collinearity to linearize the massive propagators. We have also stripped off the couplings and the prefactors arising from the expansion of the coherent state. We have furthermore defined $q_i$ as the momentum of the $i$'th emitted photon while $\tilde{q_i}$ as the sum of photon momenta emitted by the $i$'th four-point vertex. So that in the diagram above, $\tilde{q_1} = q_1 + q_2$ and $\Tilde{q}_2 = q_3 + q_4$ etc.  We then sum over permutations using the identity \footnote{See e.g.~\cite{Saotome:2012vy} for a proof.}: 
\begin{equation}
\begin{aligned}
 \hat{\delta} (\sum_{i=1}^{N} \omega_i) \left[ \frac{1}{(\omega_1 + i\varepsilon)(\omega_1 + \omega_2 +i\varepsilon) \cdots (\sum_{i=1}^{N-1} \omega_i + i\varepsilon)}\right] + \text{permutations}
 = \prod_{i=1}^{N} \hat{\delta}(\omega_i) ,
\end{aligned}
\end{equation}
where the sum runs over permutations of the vertices. Note that this is not equivalent to permuting the photon momenta, since in our notation we have labeled the sum of two momenta flowing through a 4-point vertex by one symbol $\tilde{q_i}$. This is remedied by inserting the factor ${(2m)!}/{m!}$ which accounts for the missing permutations. Attaching this prefactor to the above identity yields
\begin{equation}
    \frac{(2m)!}{m!}\left[ \prod_{j = 2m+1}^{N} \hat{\delta}(2p \cdot q_j) (\varepsilon_j \cdot p)\right] \left[\prod_{i}^{m} \hat{\delta}(2p \cdot \tilde{q_i})(\varepsilon \cdot \varepsilon^*) \right] \exp \left[ -i( p  + (\sum_{i=1}^{N} q_i)) \cdot x \right] .
\end{equation}
This is the contribution to the $n$-point amplitude of the diagrams with $r$ four-point vertices. To get the full amplitude, we have to account for the $ {N\choose 2m} $ ways of making this choice. It remains to sum over all allowed values of $m$, which takes the range $0 < m <N/2$ when $N$ is even and $0 < m <(N-1)/2$ when $N$ is odd, we will denote this value by $\lfloor N/2 \rfloor$. The $\mathcal{O}(e^N)$ contribution is
\begin{equation}
\begin{aligned}
 \mathcal{J}_{2 + N} = \sum_{m=0}^{ \lfloor N/2 \rfloor} \frac{N!}{2m! (N-2m)!}  \frac{(2m)!}{m!} \left[ \prod_{j =2m+1}^{N} \hat{\delta}(2p \cdot q_j) (\varepsilon_j \cdot p)\right] \left[\prod_{i}^{m} \hat{\delta}(2p \cdot \tilde{q_i})(\varepsilon \cdot \varepsilon^*) \right] \\
 \times \exp \left[ -i( p + (\sum q_i)) \cdot x \right]
\end{aligned}
\end{equation}
Finally, we restore the couplings along with prefactors arising from the expansion of the coherent state:
\begin{equation}
    \frac{1}{N!} (-2ie)^{N-2m}(ie^2)^{m} \left[\prod_{j}^{N} \int d \Phi (q_j) \alpha(q_j)\right] \left[\prod_{j}^{m} \int d \Phi (q_i,q'_i) \alpha(q_i) \alpha^*(q'_i)\right],
\end{equation}
which gives the result
\begin{equation}\label{factorization}
\begin{aligned}
  \mathcal{J}_{2 + N} = e^{- i p  \cdot x} & \sum_{m=0}^{\lfloor N/2 \rfloor} \frac{1}{m! (N-2m)!}   \left[ (-2ie)^{N-2m}\prod_{j}^{N -2m} \int d \Phi (q_j) e^{i q_j \cdot x}\alpha(q_j)\hat{\delta}(2p \cdot q_j) (\varepsilon_j \cdot p)\right] \\
& \times \left[(ie^2)^m\prod_{i}^{m}  \int d \Phi (q_i,q'_i) e^{i( q_i + q'_i) \cdot x}\alpha(q_i) \alpha^*(q'_i)\hat{\delta}(2p \cdot \tilde{q_i})(\varepsilon \cdot \varepsilon^*)\right].
\end{aligned}
\end{equation}
This is the total contribution of the $\mathcal{O}(e^N)$ diagrams with the LSZ reduction applied to all legs except the outgoing leg. Adding to this forward scattering contribution, and making use of the explicit form of the coherent waveshape $\alpha$ defined in \eqref{EMImpulsiveWaveshape}, we see that \eqref{factorization} coincides with the expansion of the exponential 
\begin{equation}\label{OffShellExp}
    \mathcal{J} \equiv \exp \left[ -ip \cdot x - i \int^{x^\LCm}_{0} \ud s \;  V_p(s) \right].
\end{equation}
where $V_p(s)$ is the Volkov phase \eqref{VolPhase}. The matrix elements $\mathcal{J}_{2 + N}$ can therefore be obtained at any order from the series expansion of $\mathcal{J}$. The amplitudes $\mathcal{A}_{2 + N}$ are then obtained from $\mathcal{J}_{2 + N}$ by truncating the outgoing leg with the prescription \eqref{LSZFinalLeg}. Likewise, let us define $\mathcal{A}$ by applying the same LSZ factor to $\mathcal{J}$:
\begin{equation}\label{exponentiation}
\begin{aligned}
    \mathcal{A} &=  \lim_{x_\LCm \rightarrow \infty} \bigg[\int\!\ud^3 {\bf x}\, e^{ip'\cdot x + ie \lambda(x)}  \,i \overset{\leftrightarrow}{\partial}_\LCp \bigg]  \exp \left[ -ip \cdot x - i \int^{x^\LCm}_{0} \ud s \;  V_p(s)  \right] \\
    & = 2p_+ \hat{\delta}(p_{+}^{\prime}-p_{+}) \hat{\delta}^2(p^{\prime}_\LCperp-p_\LCperp+e a_\LCperp(\infty)) \lim_{x_\LCm \rightarrow \infty} \exp \left[ i(p' -p) x^- - i \int^{x^\LCm}_{0} \ud s \;  V_p(s) \right].
\end{aligned}
\end{equation}
The final limit is evaluated by noting that, due to the deformed delta function, the mass shell condition now implies
\begin{equation}
     p'_- - p_- = V_p(\infty).
\end{equation}
Using this and integrating by parts, it is easy to check that the limit above can be written as
\begin{equation}
     \mathcal{A} = 2p_+ \hat{\delta}(p_{+}^{\prime}-p_{+}) \hat{\delta}^2(p^{\prime}_\LCperp-p_\LCperp+e a_\LCperp(\infty)) \exp \left[ i \int^{\infty}_{0} \ud s \frac{\ud }{\ud s} \;  V_p(s) \right],
\end{equation}
in agreement with the full amplitude in \eqref{QED2point}.

\bibliographystyle{JHEP}
\bibliography{ThreePointBib}

\end{document}